\documentclass[aps,prx,amsmath,amssymb,floatfix,twocolumn,superscriptaddress,nofootinbib,10pt]{revtex4-2}

\usepackage{graphicx}
\usepackage{bm}
\usepackage{stackengine}
\usepackage{amsmath} 
\usepackage{amssymb} 
\usepackage{enumitem} 
\usepackage{hyperref}
\usepackage{color} 
\usepackage{comment}
\usepackage{natbib}

\newcommand{\AAm} {{Aubry-Andr{\'e}} } 
 
\newcommand{\PBC} { {\rm pbc} }

\begin{document} 

\title{Quasiperiodicity-induced bulk localization with self similarity in non-Hermitian systems
}

\author{Yu-Peng Wang}
\affiliation{Institute of Physics, Academia Sinica, Taipei 115201, Taiwan}

\author{Chuo-Kai Chang}
\affiliation{Institute of Physics, Academia Sinica, Taipei 115201, Taiwan}

\author{Ryo Okugawa}
\affiliation{Department of Applied Physics, Tokyo University of Science, Tokyo 125-8585, Japan
}

\author{Chen-Hsuan Hsu}
\affiliation{Institute of Physics, Academia Sinica, Taipei 115201, Taiwan}

\date{\today}

\begin{abstract}

We analyze the localization behavior in a non-Hermitian system subject to a quasiperiodic onsite potential. We characterize localization transitions using multiple quantitative indicators, including inverse participation ratio (IPR), eigenstate fractal dimension (EFD), extended eigenstate ratio (EER), and spectral survival ratio. Despite the breaking of self-dual symmetry due to non-Hermiticity, our results reveal the existence of a critical potential strength, with its value increasing linearly with the nearest-neighbor antisymmetric hopping term. On the other hand, the inclusion of longer-range hopping not only enriches the topological properties but also gives rise to novel localization phenomena. In particular, it induces the emergence of mobility edges, as evidenced by both IPR and EFD, along with distinct features in the spectrum fractal dimension, which we extract using the box-counting method applied to the complex energy spectrum. Additionally, we uncover self-similar structures in various quantities, such as EER and complex eigenvalue ratio, as the potential strength varies. 
These findings highlight important aspects of localization and  fractal  phenomena in non-Hermitian quasiperiodic systems.

\end{abstract}

\maketitle

\section{introduction}

Localization phenomena have been extensively studied in low-dimensional systems driven by either disorder~\cite{Anderson:1958, Abrahams:1979, Billy:2008} or quasiperiodic~\cite{Aubry:1980, Sokoloff:1985, Roati:2008, Biddle:2010,Biddle:2011, DominguezCastro:2019} potentials.
In Anderson's seminal work~\cite{Anderson:1958}, localization arises from random disorder, leading to all states becoming localized in one- and two-dimensional systems for any nonzero disorder. 
Subsequent studies extended these investigations to systems with incommensurate or quasiperiodic potentials~\cite{Aubry:1980, Sokoloff:1985, Roati:2008, Biddle:2010}. 
When the Hamiltonian preserves its form under Fourier transformation and therefore preserves the self duality, there is a critical threshold of the quasiperiodic potential strength at which the system undergoes the localization-delocalization transition~\cite{Aubry:1980,Sokoloff:1985}.   
In systems lacking this self-dual symmetry, the threshold may be absent, leading to the emergence of mobility edges~\cite{Biddle:2010,Deng:2019,Yao:2019}. 
Moreover, quasiperiodic systems often exhibit self-similar fractal structures, as seen in the energy spectrum of the \AAm model~\cite{Yao:2019,Wu:2021}. At the critical point, the fractal (Hausdorff) dimension offers a quantitative measure for the spectral fractal structures~\cite{Tang:1986, Kohmoto:1987}. Quasiperiodicity also arises from the incommensurability through a bichromatic  potential, which reveals unusual fractal dimensions  depending on the system parameters~\cite{Yao:2019}.

Interestingly, the appearance of fractals extends beyond a purely mathematical concept, as they also affect physical properties and observables. Famous examples include Hofstadter's butterfly~\cite{Hofstadter:1976}, which arises from butterfly-like fractal energy bands under external magnetic fields, as observed in (bilayer) graphene on hBN~\cite{Dean:2013,Ponomarenko:2013}. In addition, moir{\'e}  quasicrystals can give rise to quasi-Bragg peaks and   mini-gaps in twisted bilayer WSe$_2$~\cite{Li:2024b}.

Recently, significant efforts have been made to elucidate the properties of non-Hermitian lattices~\cite{Bender:2007,Bergholtz:2021, Okuma:2023, Esaki:2011, Kunst:2018, Yao:2018,  Kawabata:2019, Okuma:2019, Yokomizo:2019, Ashida:2020, Okuma:2020,  Zhang:2020, Okugawa:2021, Wang:2023b,Nakai:2024,Shimomura:2024,Yoshida:2024,CKC:2025} 
distinct from their Hermitian counterparts. In particular, a unique localization phenomenon known as non-Hermitian skin effect can emerge even without disorder, where eigenstates localize near the system's boundaries under open boundary conditions (OBC), as exemplified by the Hatano-Nelson model~\cite{Hatano:1996, Hatano:1997}. Interestingly, this effect is linked to a topological property, where the complex energy spectrum allows for the definition of a winding number as a topological invariant characterizing the presence of skin effects~\cite{Okuma:2020, Zhang:2020,Borgnia:2020}.

Generalizations to systems exhibiting both non-Hermiticity and quasiperiodicity have uncovered unprecedented localization phenomena~\cite{Liu:2020a, Liu:2020b, Jiang:2019, Zeng:2020a, Xia:2022, Manna:2023, Acharya:2024, Li:2024, Padhi:2024, Peng:2024,  Xing:2025}.
These systems exhibit features absent in their Hermitian counterparts, including
generalizations of mobility edges in complex spectra~\cite{Liu:2020a,Liu:2020b, Zeng:2020a, Xia:2022,Gandhi:2024,Li:2024}, as well as the interplay between non-Hermitian skin effects and quasiperiodicity-induced localization~\cite{Padhi:2024, Zeng:2020a, Jiang:2019, Peng:2024, Acharya:2024}. 
While the \AAm duality has been generalized to non-Hermitian systems~\cite{Zeng:2020a,Liu:2020b}, the introduction of various tight-binding terms can break this duality and, consequently, requires the concept of mobility edges to be invoked for characterizing these systems, either numerically~\cite{Liu:2020a} or analytically~\cite{Xia:2022,Li:2024}. 
Among these additional terms, hoppings beyond the nearest-neighbor constitute straightforward additions to existing models and can be in various tight-binding forms. 
While models with exponential hopping, which preserve either parity-time symmetry~\cite{Liu:2020a} or generalized self-dual symmetry~\cite{Liu:2020b},
have been extensively studied, those with finite-range hopping terms that lack these symmetries remain less explored.
Additionally, while previous studies have examined the competition between non-Hermitian skin effects under OBC and quasiperiodicity-driven localization~\cite{Padhi:2024,Jiang:2019,Peng:2024}, the influence of longer-range hopping on quasiperiodicity-induced bulk localization under periodic boundary conditions (PBC), where skin modes are suppressed, remains unclear.
In addition to localization properties, we aim to explore how fractal structures, characteristic of quasiperiodic systems, emerge and evolve when combined with non-Hermiticity~\cite{Sun:2024}.

In this work, we investigate the generalized Hatano-Nelson model with longer-range hopping terms and a quasiperiodic onsite potential. Incorporating both additional ingredients allows us to investigate the interplay between the two. 
Despite the breaking of the self-dual symmetry, our numerics demonstrate the robustness of a critical potential strength  against non-Hermiticity when the longer-range hopping is absent. 
Utilizing the averaged eigenstate fractal dimension (EFD) as an indicator, we show that the critical potential strength increases linearly with the antisymmetric  hopping amplitude.
Applying the box-counting method to  the complex energy spectra, we deduce the spectrum fractal dimension (SFD) and find that its  minimum coincides with the critical potential strength. 
The inclusion of longer-range hopping terms not only enriches the topological phases characterized by the spectral winding number but also transforms the critical point into a regime where extended and localized states coexist.
This observation motivates the introduction of two diagnostic quantities, the extended eigenstate ratio (EER) and the spectral survival ratio, which exhibit self-similar structures as the potential strength increases, resembling the Devil's staircase~\cite{Robert:2021}.
By providing multiple quantitative indicators for both localization and fractal features, our work provides an extensive analysis on the interplay between the non-Hermiticity and quasiperiodicity in one-dimensional lattices.

The rest of the article is organized as follows. 
In Sec.~\ref{Sec:Hamiltonian}, we introduce our model and revisit its basic properties when the onsite potential is absent.
In Sec.~\ref{Sec:Localization}, we discuss the localization phenomena in our model induced by the onsite potential and introduce several quantitative indicators for characteristics, including inverse participation ratio (IPR), normalized participation ratio (NPR), and EFD. 
In Sec.~\ref{Sec:Fractal}, we discuss self-similar fractal structures present in our model.
In Sec.~\ref{Sec:SFD}, we introduce the SFD by applying the box-counting method to the complex energy spectrum. 
In Sec.~\ref{Sec:hidden}, we introduce the EER and 
discuss its self-similar features as the potential strength varies. 
In Sec.~\ref{Sec:ss}, we discuss how the self similarity arises at the critical point and close to it. 
Finally, we discuss our results and provide an outlook in Sec.~\ref{Sec:discussion}.
In Appendix~\ref{Appx:Spectral}, we review the spectral and topological properties of our model in the absence of the onsite potential.
In Appendix~\ref{Appx:r_c}, we discuss additional self-similar structure in the complex eigenvalue ratio and compare them with the EER.
In Appendix~\ref{Appx:fractal}, we provide a more detailed analysis on the self similar structure for different parameter regimes.

\section{Hamiltonian}
\label{Sec:Hamiltonian}

\begin{figure}[h]
    \centering
    \includegraphics[width=0.47\textwidth]{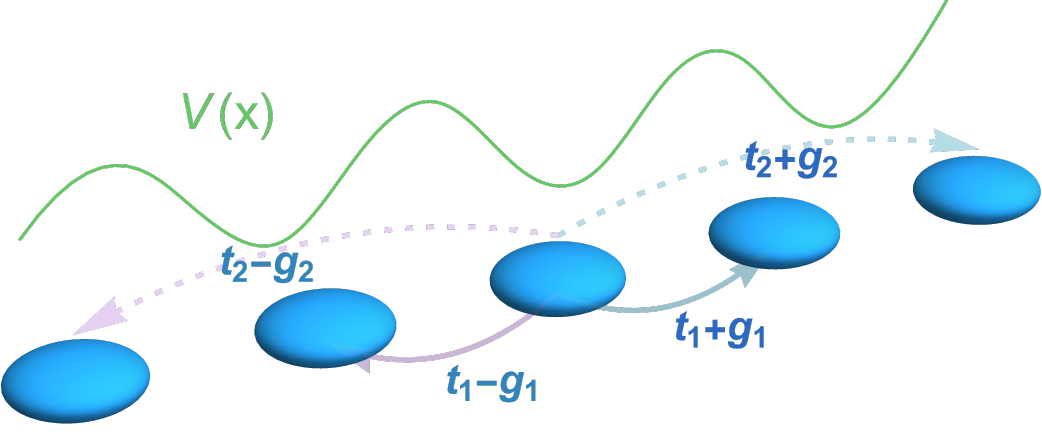}
    \caption{Schematic setup of a non-Hermitian lattice subject to an onsite potential, as described by Eq.~\eqref{Eq:H_nH+qp}. The arrows represent asymmetric hoppings  $(t_{1,2} \pm  g_{1,2})$ to the right/left directions, and the wavy curves denote a real onsite potential  $V(x) = V_0 \cos ( 2\pi x / \lambda_0 )$ with the wavelength $\lambda_0$. 
     }
    \label{Fig:setup}
\end{figure}

We explore the spectral and transport properties of a one-dimensional lattice  with asymmetric hopping terms and a quasiperiodic onsite potential. We sketch our setup in Fig.~\ref{Fig:setup} for illustration and describe them as 
\begin{subequations}
    \label{Eq:H_nH+qp}
\begin{eqnarray}
  H_{\rm  } & = & H_{\rm nH } + V_{\rm qp} ,  
\end{eqnarray}
with 
\begin{eqnarray}
  H_{\rm nH} & = &
    - \sum_{n =1}^{n_{\rm max}} \sum_{j=1}^{N}  \Bigg[ (t_n +  g_n ) c^{\dagger}_{j+n} c_{j} + (t_n -  g_n ) c^{\dagger}_{j-n} c_{j} \Bigg] , \nonumber \\
    && 
    \label{Eq:H_nH}
    \\
  V_{\rm qp}  &=&  \sum_{j=1}^{N} 
    V_0 \cos \Bigg( \frac{2\pi a_0}{\lambda_0} j \Bigg) c^{\dagger}_j c_{j}
  ,      
   \label{Eq:V_qp}
\end{eqnarray}
\end{subequations}
where $c_{j }^{\dagger}$ ($c_{j }$) represents the creation (annihilation) operator of a spinless  particle on the $j$th site in a one-dimensional lattice of $N$ sites. The real parameters $t_n$ ($g_n$) denote the symmetric (antisymmetric) components of the $n$th-nearest-neighbor hopping terms. 
In the above, $V_0$ is the potential strength, $a_0$ is the lattice constant, and $\lambda_0$ is the wavelength of the onsite potential, which controls the (quasi)periodicity with commensurability determined by the ratio $\lambda_0 / a_0$. 
Throughout this work, we consider an irrational ratio of $\lambda_0 / a_0 = (\sqrt{5} + 1)/2$, such that quasiperiodicity arises from the incommensurate potential landscape relative to the lattice\footnote{Alternatively, one can also utilize approximate quasiperiodicity using the Fibonacci sequence, which preserves the parity-time symmetry in certain models~\cite{Xia:2022} and can be more straightforwardly achieved in experimental setups. 
} and take $t_1=1$ as an overall energy scale, unless otherwise stated.

 \begin{figure}[t]
    \centering
    \stackinset{l}{0pt}{b}{2.67cm}{\colorbox{white}{(a)}}
    {\includegraphics[width=0.48\linewidth]{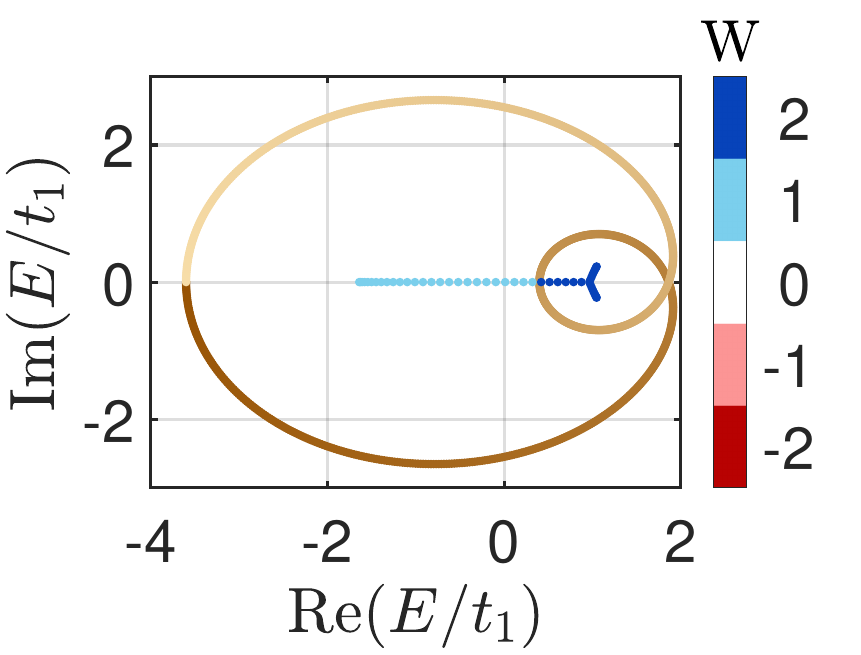}}
    \stackinset{l}{-9pt}{b}{2.67cm}{\colorbox{white}{(b)}}
    {\includegraphics[width=0.485\linewidth]{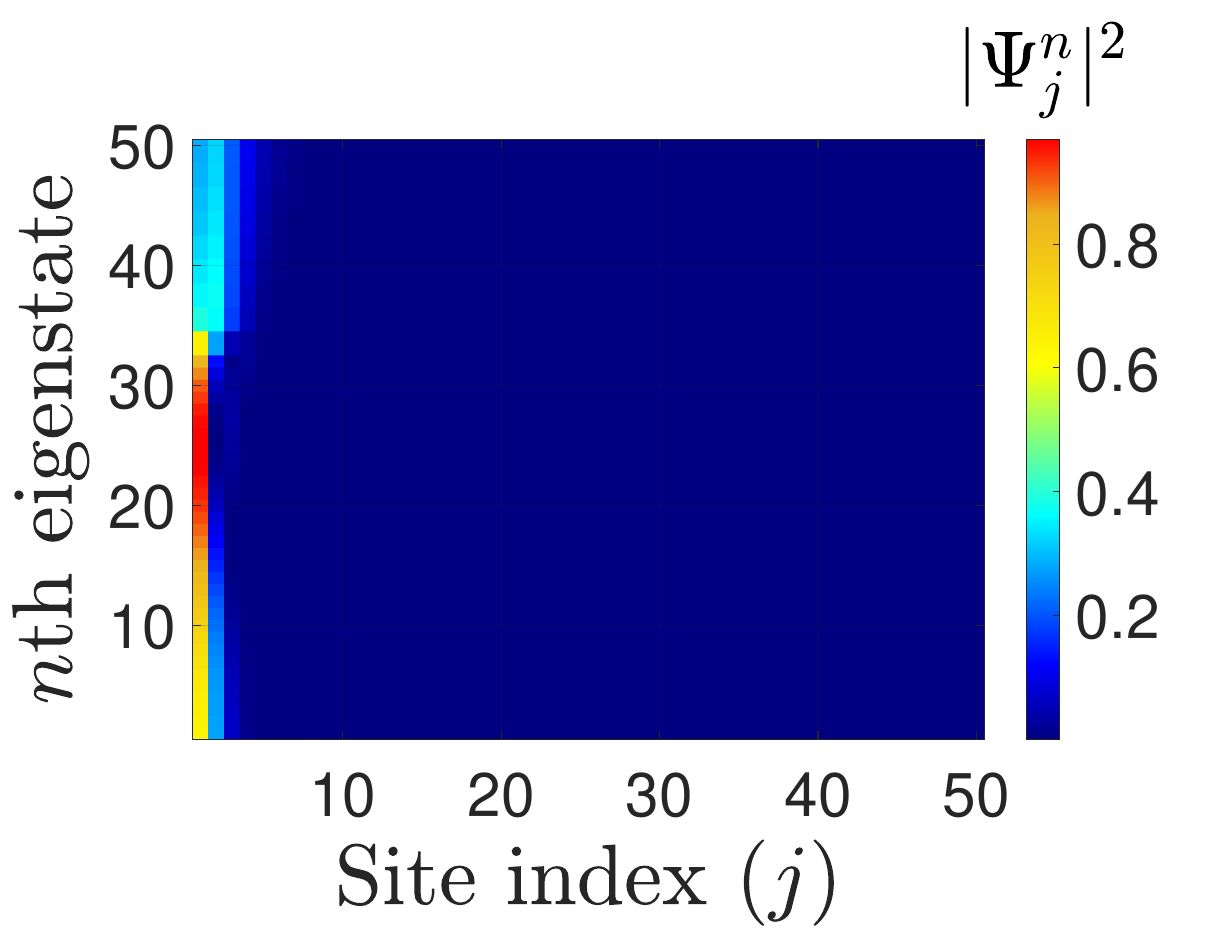}}\\
    \caption{(a) PBC (curve) and OBC (dot) spectra of Eq.~\eqref{Eq:H_nH+qp} for $V_0 = 0$ and $N=50$. 
    The PBC spectrum represents a function of momentum $k$, ranging from $0$ (brown) to $2\pi/a_0$ (light brown), while the color of the OBC spectrum indicates the winding number $W(E_r)$ defined in Eq.~\eqref{Eq:Winding}, with the reference point $E_r$ set to the corresponding OBC eigenvalues. 
    (b) Spatial distribution of all the OBC (right) eigenstates.
    The values of the adopted parameters  are given by $(g_1, t_2, g_2) = (-0.7, 0.8, -0.8)$ and $t_{n \ge 3} = g_{n \ge 3} = 0$.  
    }
    \label{Fig:nh}
\end{figure}

In Eq.~\eqref{Eq:H_nH+qp}, we incorporate longer-range hopping ($n > 1$,  up to $n = n_{\rm max}$), which has been shown to result in a rich phase diagram with higher winding numbers in Hermitian lattices~\cite{Niu:2012}. Accordingly, we anticipate enriched properties in our non-Hermitian settings.
To demonstrate the effects of hopping beyond the nearest-neighbor sites, we will incorporate the next-nearest-neighbor term, that is, $n_{\rm max} =2$ in Eq.~\eqref{Eq:H_nH}; even longer-range terms can be added straightforwardly. 
Below we briefly review the properties of the hopping terms $H_{\rm nH}$ in Eq.~\eqref{Eq:H_nH}, before investigating the localization driven by $ V_{\rm qp}$ in Eq.~\eqref{Eq:V_qp} in Sec.~\ref{Sec:Localization}--\ref{Sec:Fractal}.

\begin{figure}[h]
    \hspace{-10pt}
    \stackinset{l}{-0.05cm}{b}{2.5cm}{\colorbox{white}{(a)}}{
    \includegraphics[width=0.45\linewidth]{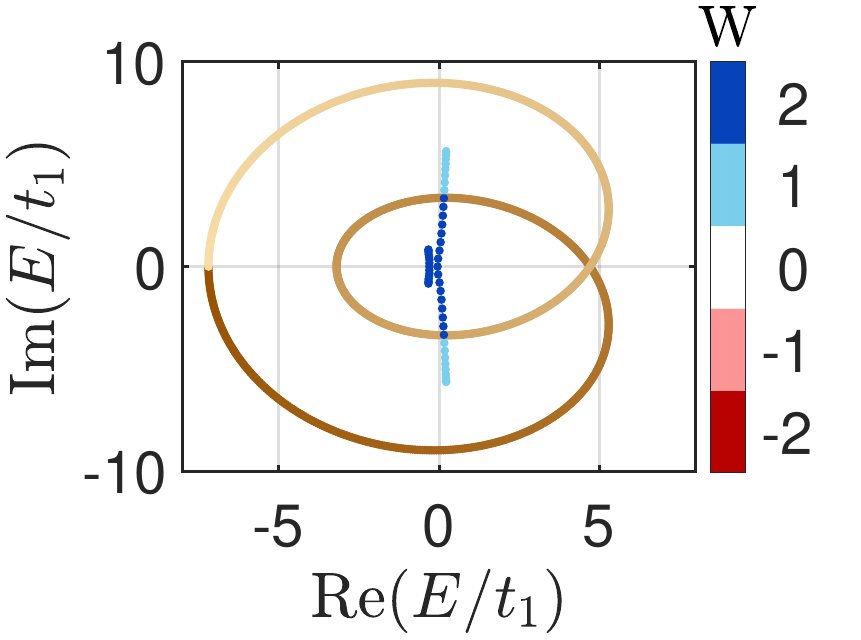}
    }
    \stackinset{l}{-0.05cm}{b}{2.5cm}{\colorbox{white}{(b)}}{
    \includegraphics[width=0.45\linewidth]{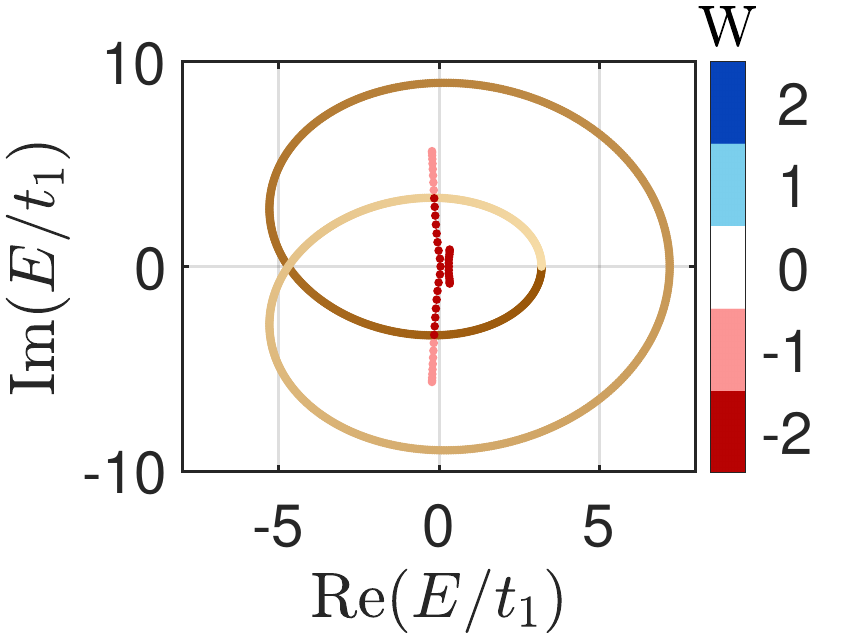}
    } \\
    \stackinset{l}{-0.05cm}{b}{2.5cm}{\colorbox{white}{(c)}}{
    \includegraphics[width=0.456\linewidth]{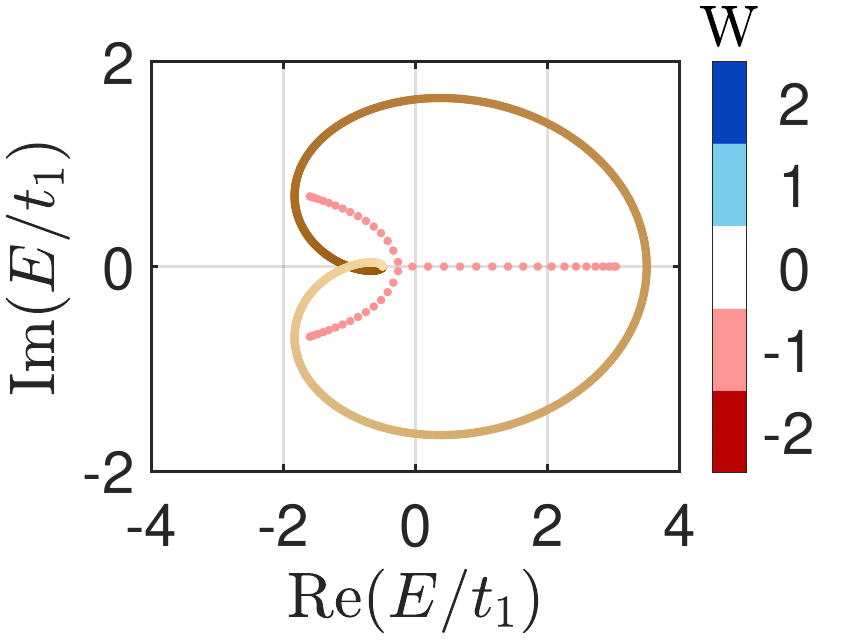}
    }
    \stackinset{l}{-0.1cm}{b}{2.5cm}{\colorbox{white}{(d)}}{
    \includegraphics[width=0.45\linewidth]{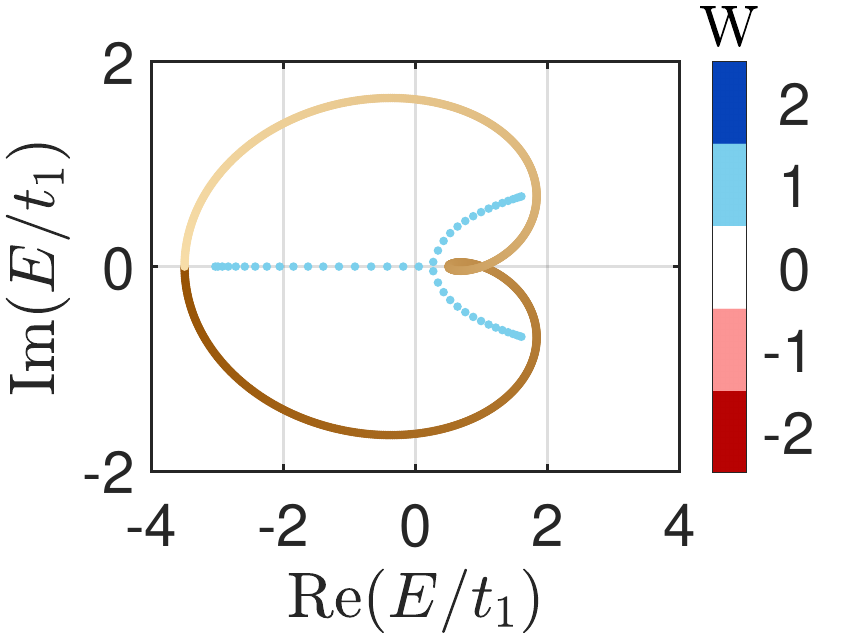}
    } \\
    \stackinset{l}{-0.05cm }{b}{2.5cm}{\colorbox{white}{(e)}}{
    \includegraphics[width=0.45\linewidth]{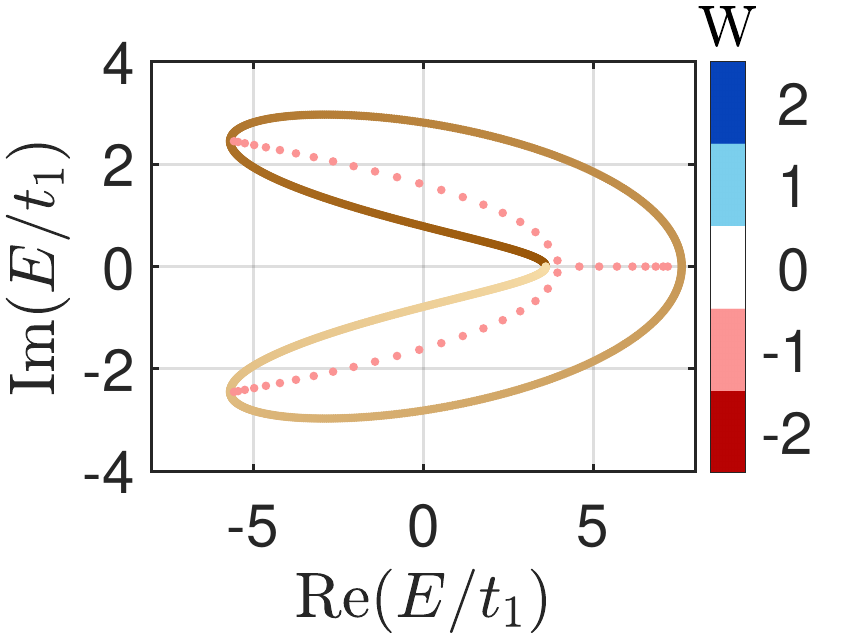}
    }
    \caption{PBC (curve) and OBC (dot) spectra of Eq.~\eqref{Eq:H_nH}; the latter are plotted with $N=50$. 
    The PBC spectra are obtained from Eq.~\eqref{Eq:E_HNL_PBC}, with 
$ k$ going from $0$ (brown) to $2\pi/a_0$ (light brown). 
    The color of the OBC spectra indicates the winding number $W(E_r)$ given by Eq.~\eqref{Eq:Winding}, with the reference point $E_r$ set to the corresponding eigenvalues.
    From Panel~(a) to (e), the parameters $(g_1 , t_2 , g_2) $ are given by  
     $(-2, 2.6, -3 )$,   $(2, -2.6, -3)$, $(0.6,  -0.75, -0.35)$, $(-0.6, 0.75,  -0.35)$, and $( 1.3,  -2.8, -0.4)$, respectively. 
    }
    \label{Fig:H_HNL_Spectra}
\end{figure}

Under PBC, the energy spectrum of $H_{\rm nH}$ forms closed loops in the complex energy plane, enabling the definition of a spectral winding number. It quantifies how many times the spectral trajectory encircles a reference point $E_r$ in the complex plane~\cite{Okuma:2020} and can be computed as
\begin{equation}
    W(E_r) = \int_0^{2\pi/a_0} \frac{dk}{2\pi i} \frac{d}{dk} \ln \left\{ \det \big[ H_{\rm nH}^{\PBC}(k) - E_r \big] \right\},
    \label{Eq:Winding} 
\end{equation}
where $H_{\rm nH}^{\PBC}(k)$ is the Bloch Hamiltonian derived from Eq.~\eqref{Eq:H_nH}.  
Examples of the spectral winding number for several parameter sets are presented in Figs.~\ref{Fig:nh}--\ref{Fig:H_HNL_Spectra}. 
As expected, the next-nearest-neighbor hopping terms make the energy spectrum more complicated than ellipses. In consequence, Fig.~\ref{Fig:nh}(a) illustrates the emergence of not only a unity winding number $W = 1$ but also a higher winding number $W = 2$.  
We also obtain the PBC spectra for a range of parameter values  and present them in Fig.~\ref{Fig:H_HNL_Spectra}. The finding that the longer-range hoppings give rise to higher winding numbers is consistent with Ref.~\cite{RafiUlIslam:2024}. 
Notably, this mirrors a similar behavior as in Hermitian systems~\cite{Niu:2012}, where longer-range terms in transverse-field quantum spin chains lead to higher winding numbers and thus multiple Majorana zero modes at the edges of open chains. 

In addition to the PBC setting, we briefly discuss the behavior of skin modes under OBC. To this end, we obtain the OBC spectra shown in Figs.~\ref{Fig:nh}--\ref{Fig:H_HNL_Spectra}, alongside the PBC spectra computed using the same parameter set and the corresponding spatial distribution of the right eigenstates [see Fig.~\ref{Fig:nh}(b)].
A distinctive characteristic of non-Hermitian Hamiltonians is that their eigenvalues and eigenstates depend on boundary conditions. 
In contrast to PBC, the density profile of the eigenstates under OBC becomes localized at one of the edges, forming skin modes; see also
Fig.~\ref{Fig:nh2} in Appendix~\ref{Appx:Spectral} for the spatial profile for selected eigenstates. Here we illustrate the profiles of the right eigenstates, while the corresponding left eigenstates localize at the opposite edge (data not shown for brevity). 

Having reviewed the properties of $H_{\rm nH}$ and introduced the spectral winding number in the PBC spectra, in the following section we investigate how these properties are influenced by the presence of $ V_{\rm qp}$ in Eq.~\eqref{Eq:V_qp}.

\section{Localization induced by the quasiperiodic onsite potential }
\label{Sec:Localization}

In this section we discuss the localization induced by the onsite potential. The non-Hermiticity due to the antisymmetric hopping terms induces skin modes that can compete with bulk localization induced by the onsite potential. To isolate the latter effect, we impose PBC to suppress the formation of skin modes and  examine spectral characteristics, spatial distributions of  eigenstate density, and localization-delocalization transitions as system parameters vary.

To proceed, we employ the concepts of the IPR and NPR, both closely associated with localization behavior. To determine whether an eigenstate is localized, the IPR and NPR are defined for the $n$th right eigenstate $|\Psi^n\rangle$, with components $\Psi^n_j$ on the $j$th site.
For a general Hamiltonian $H$, the IPR and its average are defined as~\cite{Park:2016, Yao:2019}
\begin{subequations}
    \begin{eqnarray} \label{Eq:IPR}
    \text{IPR}_n &\equiv& \text{IPR}(|\Psi^n\rangle) \equiv \frac{\sum_{j=1}^N |\Psi_j^n|^4}{(\langle \Psi^n | \Psi^n \rangle)^2}, \\
    \langle \text{IPR} \rangle_H &\equiv& \frac{1}{N} \sum_{n=1}^N \text{IPR}_n,
    \label{Eq:IPR_H}
\end{eqnarray}
\end{subequations}
whereas the NPR  and its average are defined as~\cite{Padhi:2024}
\begin{subequations}
\begin{eqnarray}
       \text{NPR}_n &\equiv& \text{NPR}(\lvert \Psi^{n} \rangle) = (N \times \text{IPR}(\lvert \Psi^{n} \rangle)^{-1} , \\
    \langle \text{NPR} \rangle_H 
    &\equiv& \frac{1}{N} \sum_{n=1}^N \text{NPR}_n.
    \label{Eq:NPR}
\end{eqnarray}
\end{subequations}
To simultaneously characterize the eigenvalues and eigenstates in our system, we combine the PBC spectra and IPR within single plots; we note that, in contrast to our approach,  Ref.~\cite{Padhi:2024} studied a similar system under the OBC, where skin modes are present and might therefore mask the quasiperiodicity effects.

\begin{figure*}[t]
    \centering
    \stackinset{l}{0cm}{b}{4.1cm}{\colorbox{white}{(a)}}{
        \includegraphics[width=0.313\textwidth]{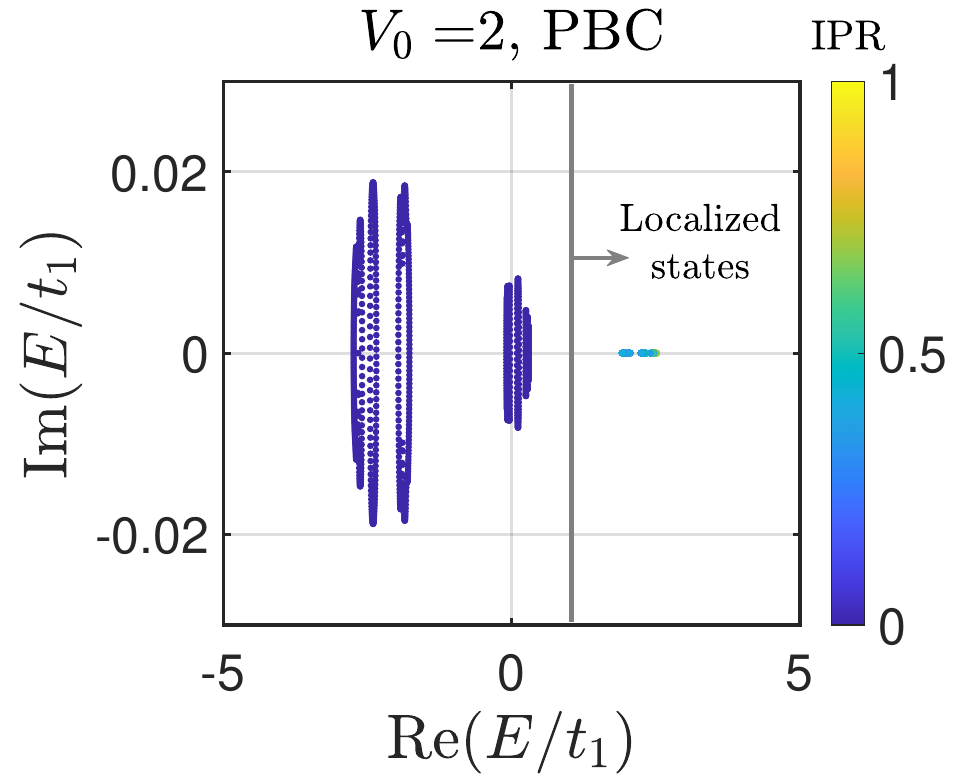}
    }
    \stackinset{l}{0cm}{b}{4.1cm}{\colorbox{white}{(b)}}{
        \includegraphics[width=0.313\textwidth]{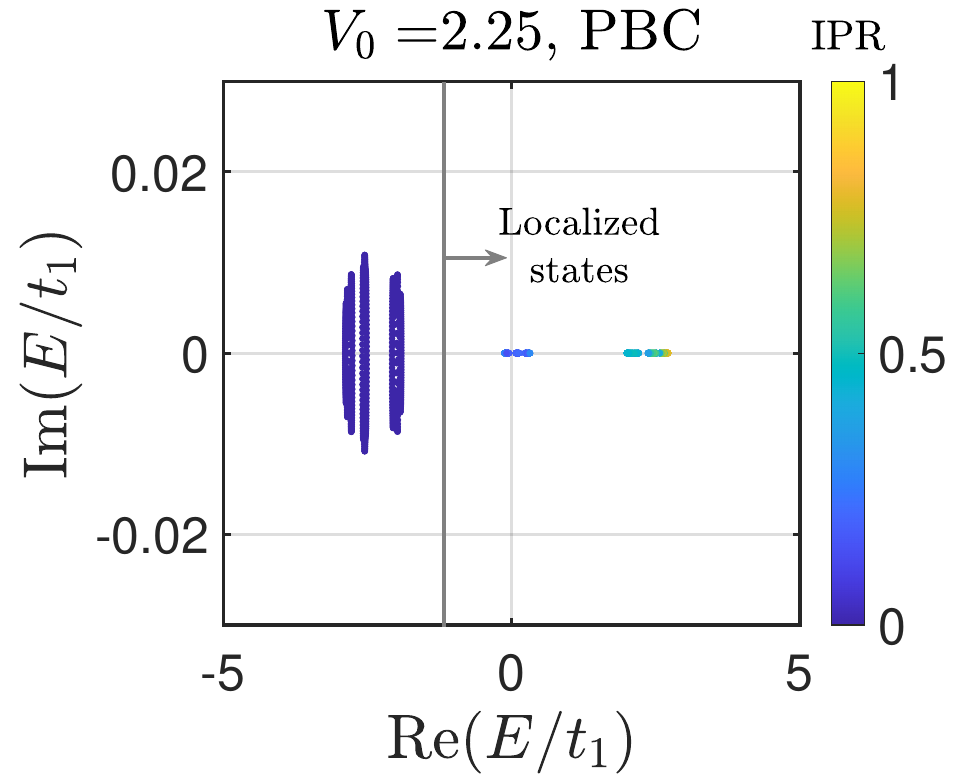}
    }
    \stackinset{l}{0cm}{b}{4.1cm}{\colorbox{white}{(c)}}{
        \includegraphics[width=0.315\textwidth]{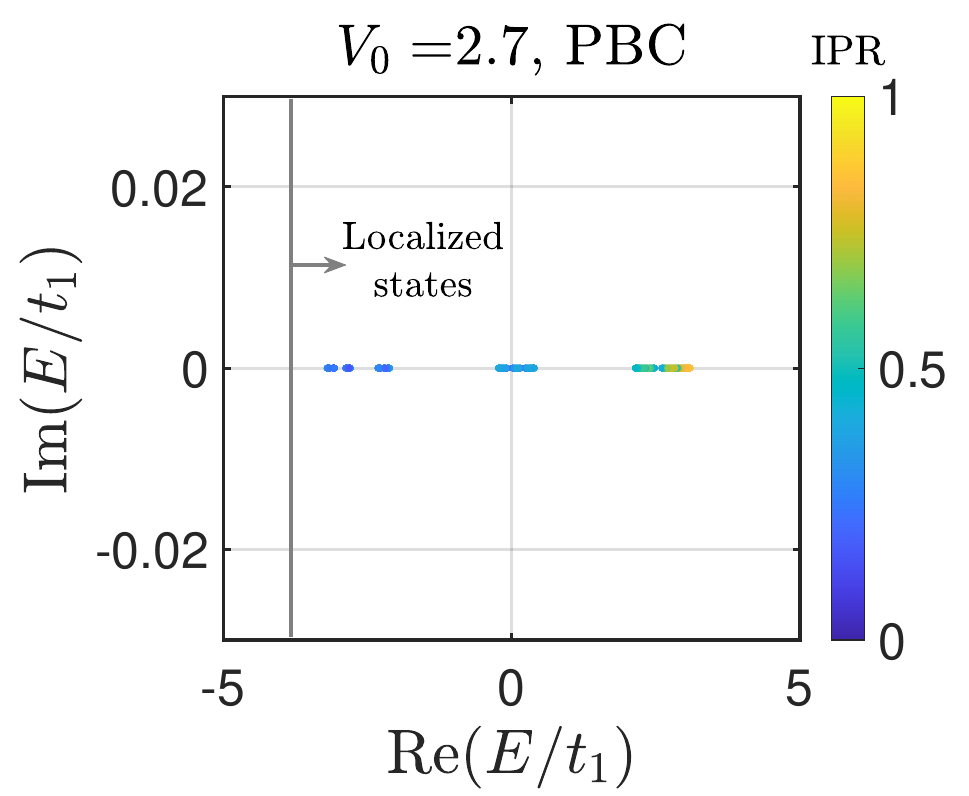}
    } \\
    \stackinset{l}{0cm}{b}{3.8cm}{\colorbox{white}{(d)}}{
    	\includegraphics[width=0.3175\textwidth]{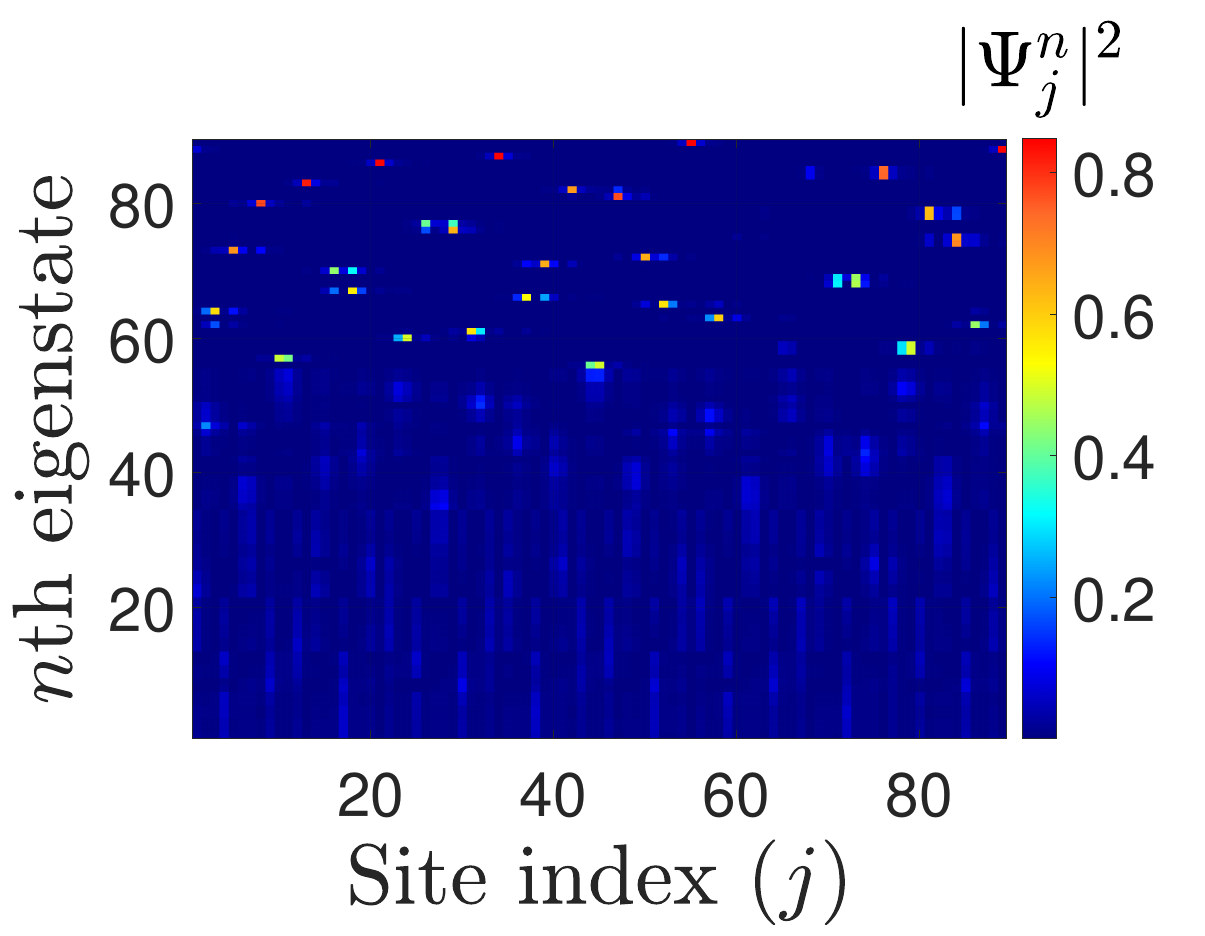}
    }
    \stackinset{l}{0cm}{b}{3.8cm}{\colorbox{white}{(e)}}{
    	\includegraphics[width=0.3175\textwidth]{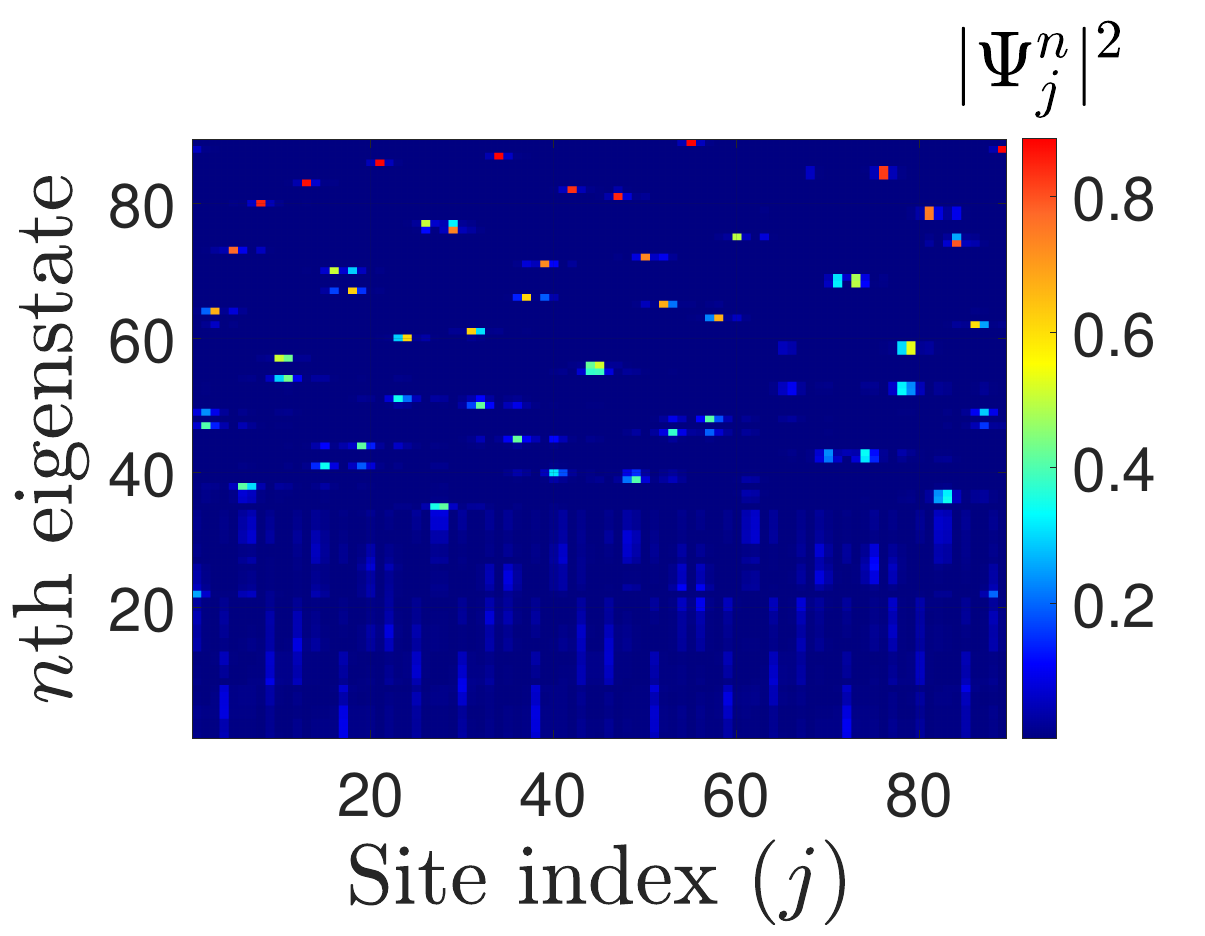}
    }
    \stackinset{l}{0cm}{b}{3.8cm}{\colorbox{white}{(f)}}{
    	\includegraphics[width=0.3175\textwidth]{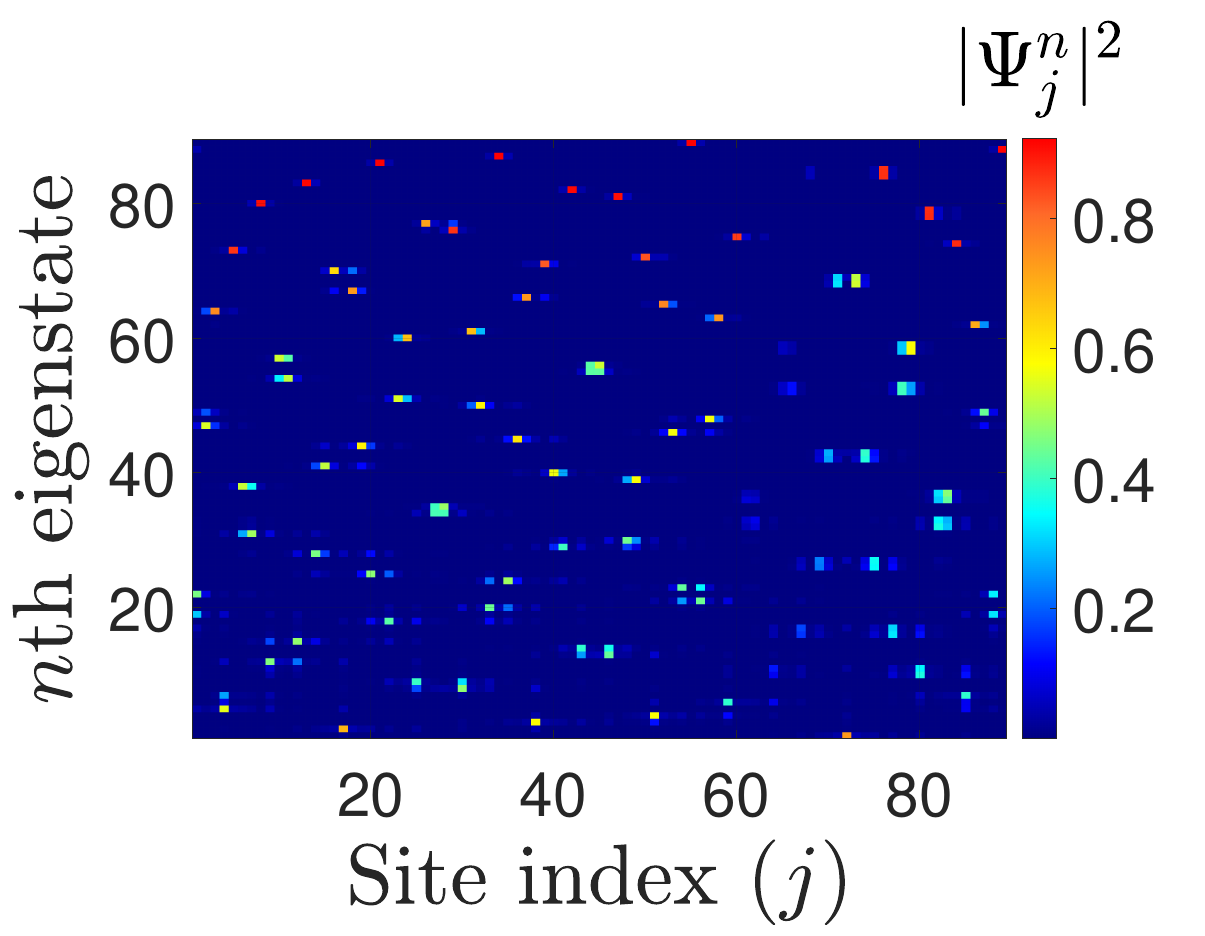}
    	}
    \caption{(a)--(c) PBC spectra and IPR of Eq.~\eqref{Eq:H_nH+qp} for $N=1597$ and
    (a) $V_0 = 2$, (b) $V_0 = 2.25$, and (c) $V_0 = 2.7$. 
    The color of the dots represents the IPR computed from Eq.~\eqref{Eq:IPR}: 
    yellow (blue) color indicates more localized (extended) states. 
    (d)--(f) Spatial profile of the right eigenstates corresponding to (a)--(c), but for $N=89$ and $\lambda_0/a_0=144/89$. The eigenstates are ordered by their eigenvalues, with a larger $n$ labeling a larger Re$(E_n)$.
    The adopted values of the other parameters are given by $(g_1,t_2,g_2) = (0.05,0.1,0)$. 
    } 
    \label{Fig:PBC-Spectra_HNL-qp} 
\end{figure*}

In addition to the IPR and NPR, we also define the EFD for the $n$th eigenstate and its average over the entire system~\cite{Deng:2019,Xia:2022},   
\begin{equation}
    \Gamma_{n} \equiv - \lim_{N \rightarrow \infty} \frac{\ln(\text{IPR}_n)}{\ln N}, 
    \quad
    \langle \Gamma \rangle_H \equiv \frac{1}{N} \sum_{n=1}^N \Gamma_n ,
    \label{Eq:EFD}
\end{equation} 
where, in practice, we increase $N$ in our numerics until convergence is achieved. 
For sufficiently large systems, we have the tendency of $\text{IPR}_n \to 0$ and $\Gamma_{n} \to 1$ for more extended states, whereas  $\text{NPR}_n \to 0$ and $\Gamma_{n} \to 0$ for more localized states,
making them effective tools to identify the energy scale or potential strength separating localized and extended states. 

In Fig.~\ref{Fig:PBC-Spectra_HNL-qp} we present the PBC spectra in the complex energy plane with IPR coloring to track localization transitions. Increasing $V_0$ from Fig.~\ref{Fig:PBC-Spectra_HNL-qp}(a) to Fig.~\ref{Fig:PBC-Spectra_HNL-qp}(b)  leads to line gaps and the formation of smaller loops, accompanied by enhanced localization and an increased number of states with real eigenvalues. 
During the process, the imaginary parts of the eigenvalues with larger real parts start to reduce and then vanish, at which the corresponding eigenstates become localized. 
Eventually, one arrives at Fig.~\ref{Fig:PBC-Spectra_HNL-qp}(c), where all the eigenvalues become real with all the states localized.
Interestingly, this evolution from complex to real spectra is marked by the disappearance of local spectral winding numbers. In consequence, there appears to be a correlation between vanishing local spectral winding numbers in the PBC spectrum and localization of the corresponding eigenstates.

In addition to the localization properties indicated by the IPR, we also look into the spatial distribution of the eigenstates. 
Consistent with Fig.~\ref{Fig:PBC-Spectra_HNL-qp}(a), where a low IPR$_n$ suggests the predominance of extended states, we observe more dispersed distributions in Fig.~\ref{Fig:PBC-Spectra_HNL-qp}(d). 
Conversely, with an increase in the $V_0$ value, there is a more pronounced increase in the number of localized states in Fig.~\ref{Fig:PBC-Spectra_HNL-qp}(e,f), aligning with the higher IPR$_n$ values in Fig.~\ref{Fig:PBC-Spectra_HNL-qp}(b,c). Note that in Fig.~\ref{Fig:PBC-Spectra_HNL-qp}(d)--(f), a smaller system size $N$  with approximate quasiperiodicity 
is chosen so that the localization phenomena are better showcased.

\begin{figure}[t]
    \centering
        \stackinset{l}{0cm}{b}{3cm}{\colorbox{white}{(a)}}{
    \includegraphics[width=0.465\linewidth]{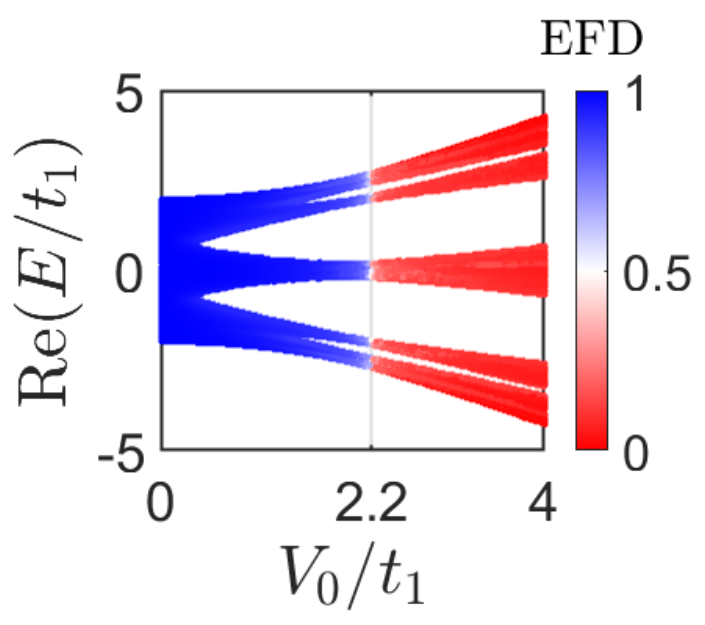}
    }
    \stackinset{l}{0cm}{b}{3cm}
    {\colorbox{white}{(b)}}{
    \includegraphics[width=0.465\linewidth]{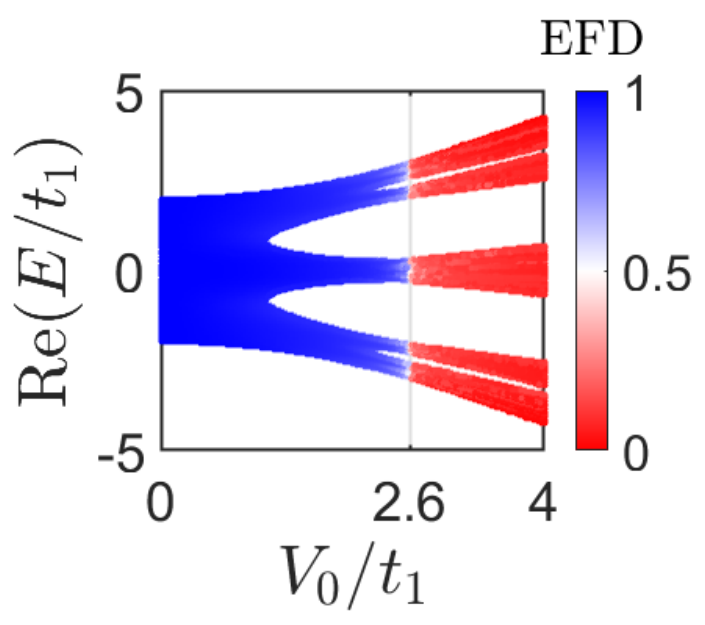}
    }
    \raggedright
    \stackinset{l}{0.1cm}{b}{2.8cm}
    {\colorbox{white}{(c)}}{
        \includegraphics[width=0.47\linewidth]{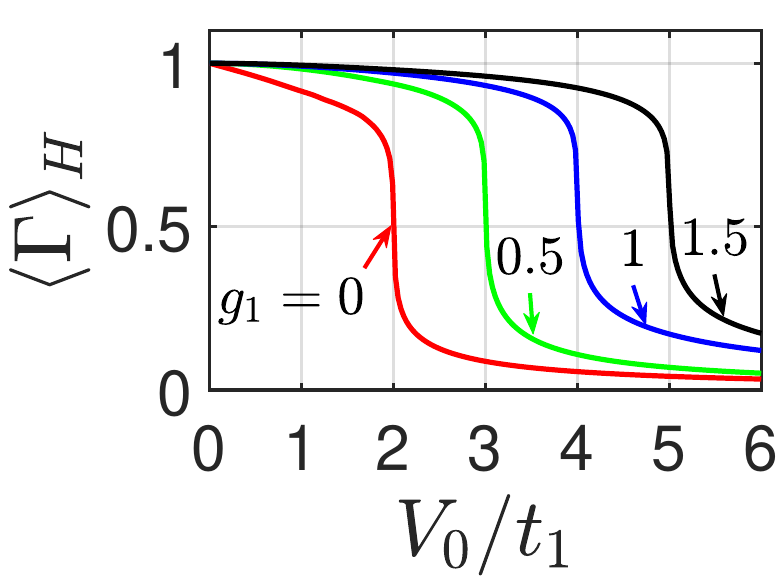}
    }
     \stackinset{l}{0.1cm}{b}{3cm}{\colorbox{white}{(d)}}{
        \includegraphics[width=0.44\linewidth]{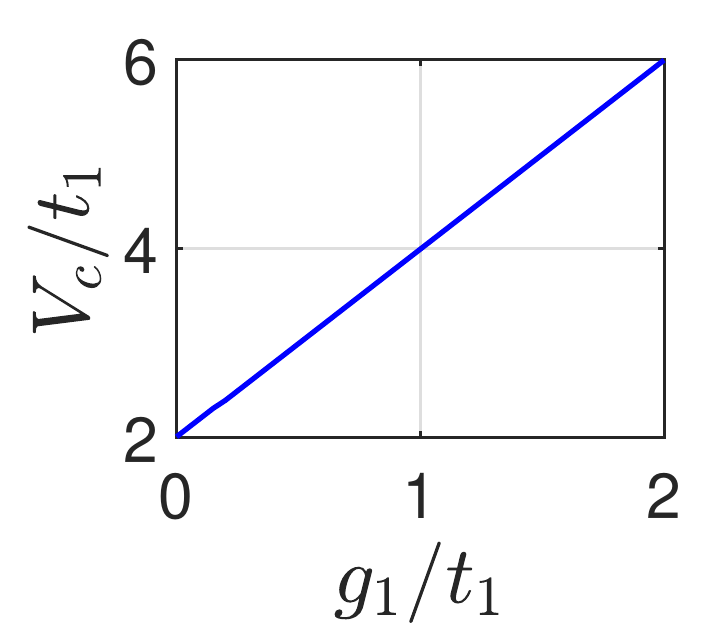}
    }
    \caption{EFD ($\Gamma_n$) and averaged EFD ($\langle \Gamma \rangle_H$) computed from Eq.~\eqref{Eq:EFD} and the deduced critical potential strength. (a) $\Gamma_n$ for $g_1 = 0.1$. (b) Similar to (a) but for $g_1 = 0.3$. 
    (c) $\langle \Gamma \rangle_H$ as a function of $V_0$ for various $g_1$ values.
    (d) Antisymmetric hopping amplitude ($g_1$) dependence of the critical potential strength ($V_c$), deduced from the $V_0$  value at which $ \langle \Gamma \rangle_H$   drops below 0.5. The adopted values of the other parameters are given by $(t_2,g_2) = (0,0)$ and $N=1597$. 
  }
  \label{Fig:EFD-vs-V0-g1}
\end{figure}

To further characterize the localization behavior, we compute the EFD in Eq.~\eqref{Eq:EFD} as a function of the potential strength $V_0$ and 
identify the localized states and localized phase for $\Gamma_{n} \to 0 $ and $\langle \Gamma \rangle_H \to 0$, respectively.
Interestingly, while our model generally lacks self-dual symmetry due to $g_1 \neq 0$, our numerical results reveal a critical potential strength $V_c$ that separates a phase consisting entirely of localized states from one containing only extended states. 
An example is shown in Fig.~\ref{Fig:EFD-vs-V0-g1}, where the longer-range hopping parameters are set to zero while $g_1$ remains nonzero, demonstrating the robustness of $V_c$ against non-Hermiticity.  
In Fig.~\ref{Fig:EFD-vs-V0-g1}(a)--(b), the states are either localized (red region) or extended (blue region), separated by a critical potential $V_c$ that depends on the model parameters, specifically $g_1$ in this case. Figures~\ref{Fig:EFD-vs-V0-g1}(c)--(d) further illustrate the transition between extended and localized states, allowing us to determine $V_c$ as the value of $V_0$ at which $\langle \Gamma \rangle_H$ drops below 0.5.  
We remark, however, that when $\langle \Gamma \rangle_H$ remains close to 0.5, the corresponding states cannot be regarded as truly extended or localized. Our use of this threshold is therefore intended only as a heuristic criterion for numerical analysis, rather than a strict definition.  
As shown in Fig.~\ref{Fig:EFD-vs-V0-g1}(d), $V_c$ exhibits a linear dependence on $g_1$, which can be empirically described as $V_c = 2 (t_1 + g_1)$ in the regime $0 \leq g_1 \leq t_1$.
We remark that, our results are consistent with the findings in Ref.~\cite{Longhi:2021}. 
However, our determination of $V_c$ follows an approach distinct from Ref.~\cite{Longhi:2021}, which focused on dynamical properties; here we identify $V_c$  using static eigenstate characteristics (EFD and IPR) under PBC.
We note that Refs.~\cite{Liu:2021c,Liu:2020b} also demonstrated the robustness of the critical potential strength against asymmetric hopping, though with different forms of hopping amplitudes in their model.

The existence of $V_c$ can be disrupted by hopping beyond the nearest-neighbor sites. Nonetheless, for a given value of $V_0$, one can still identify an energy scale that separates localized and extended states, as suggested in Fig.~\ref{Fig:PBC-Spectra_HNL-qp}.
Specifically, we find that states with real eigenvalues tend to have an IPR significantly larger than zero, indicating their localized nature. For a given potential strength, there exists a specific window in the real part of the energy within which states remain localized, as highlighted by the gray arrows in Fig.~\ref{Fig:PBC-Spectra_HNL-qp}(a)--(c).

\begin{figure}[t]
    \centering    
    \stackinset{l}{0.1cm}{b}{3.cm}{\colorbox{white}{(a)}}{
        \includegraphics[width=0.465\linewidth]{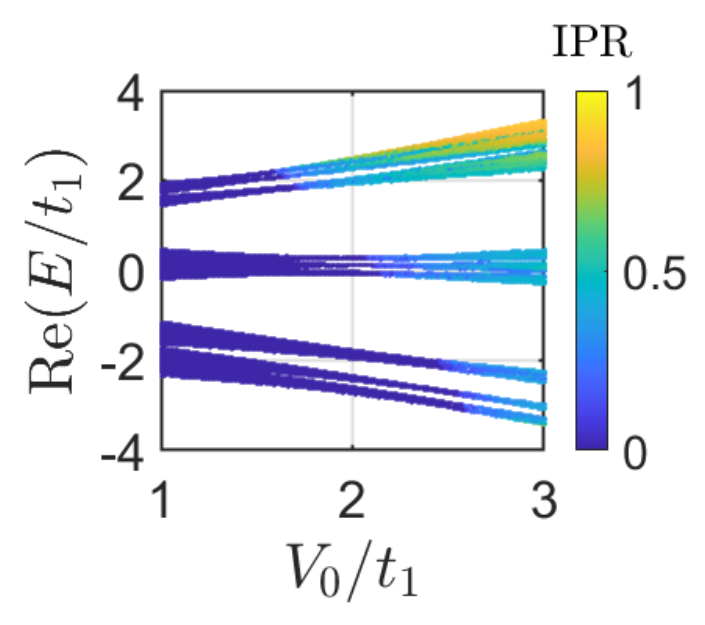}
        }
    \stackinset{l}{0.1cm}{b}{3.cm}{\colorbox{white}{(b)}}{
        \includegraphics[width=0.465\linewidth]{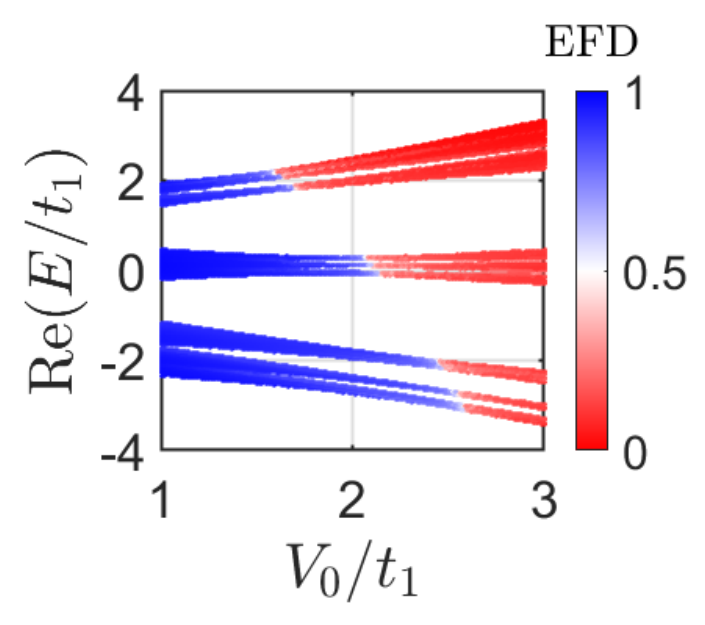}
    }    
    \caption{(a) IPR$_n$ computed from Eq.~\eqref{Eq:IPR} and (b) EFD ($\Gamma_n$) computed from Eq.~\eqref{Eq:EFD}  for the parameter set of $(g_1,t_2,g_2) = (0.05,0.1,0)$ and $N=1597$.
   }
    \label{Fig:EFD}
\end{figure}

To track how this behavior evolves with the onsite potential strength $V_0$, we combine the IPR and the real part of the PBC spectra in Fig.~\ref{Fig:EFD}(a). Here, the real part of the eigenvalues is utilized to identify the energy scale separating localized and extended states, thereby generalizing the concept of mobility edges from traditional Hermitian systems.
Accompanying the IPR, we plot the EFD in Fig.~\ref{Fig:EFD}(b), which exhibits a sharp transition from localized to extended states for finite $g_1$, indicating a well-defined generalized mobility edge. 
Notably, the next-nearest-neighbor hopping terms not only influence the winding number (see Fig.~\ref{Fig:nh}) but also impact transport properties. By comparing Fig.~\ref{Fig:EFD}(b) with Fig.~\ref{Fig:EFD-vs-V0-g1}, we observe that a finite $t_2$ induces a generalized mobility edge, consistent with the behavior known in Hermitian systems~\cite{Biddle:2011}.
Additionally, a closer look at around $V_0/t_1 \approx 1.5$ and Re($E/t_1) \approx 2 $ in Fig.~\ref{Fig:EFD}(b) shows that the mobility edge varies with $V_0$ in a non-linear manner.
  
Before concluding this section, we note that with multiple quantities at our disposal, we can evaluate the most effective indicators for mobility edges. Specifically, the IPR$_n$ defined in Eq.~\eqref{Eq:IPR} and the EFD ($\Gamma_n$) in Eq.~\eqref{Eq:EFD} both serve to deduce the mobility edges between localized and extended states.
By comparing Fig.~\ref{Fig:EFD}(a) with Fig.~\ref{Fig:EFD}(b), it becomes evident that $\Gamma_n$ exhibits a consistently sharper transition than IPR$_n$, making it a more effective tool for identifying critical transitions in the system. In addition to localization phenomena, the quasiperiodic potential can also give rise to fractal or self-similar features, which we explore next.

\section{Fractal and self-similar features  }
\label{Sec:Fractal}

In this section we discuss the fractal  or self-similar features in our model. To this end, we first compute the fractal dimension from the complex energy spectra and then discuss self-similar structures appearing in various quantities characterizing the localization phenomena.  

\subsection{Spectrum fractal dimension (SFD) }
\label{Sec:SFD}

In this section, we further characterize the spectral structures by extending the box-counting method~\cite{Theiler:1990} to the two-dimensional complex energy plane\footnote{It should be noted that this numerical method might only provide an upper bound of the true fractal dimension~\cite{Theiler:1990}.
}.
To distinguish from the EFD introduced in Sec.~\ref{Sec:Localization}, here we define the SFD as the fractal dimension of the PBC energy spectrum. 
While the SFD has been studied in the literature on Hermitian systems~\cite{Tang:1986,Kohmoto:1987,Yao:2019}, here our emphasis is on the non-Hermitian regime with complex energy spectrum, which makes the box-counting procedure more involved.

\begin{figure}[t]
    \centering
    \stackinset{l}{-0.1cm}{b}{2.55cm}{\colorbox{white}{(a)}}{
        \includegraphics[width=0.46\linewidth]{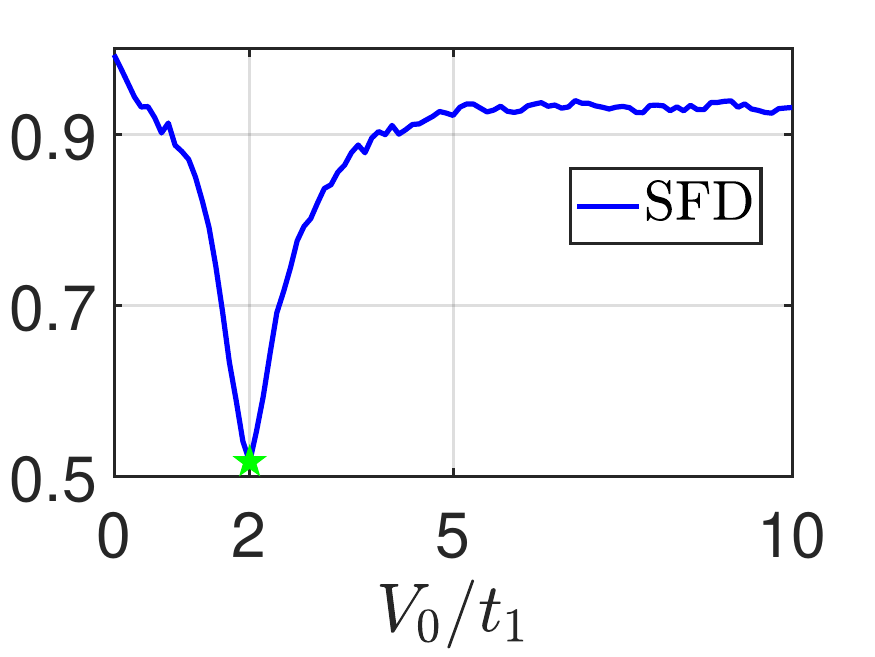}
    }
    \stackinset{l}{-0.1cm}{b}{2.55cm}{\colorbox{white}{(b)}}{
        \includegraphics[width=0.46\linewidth]{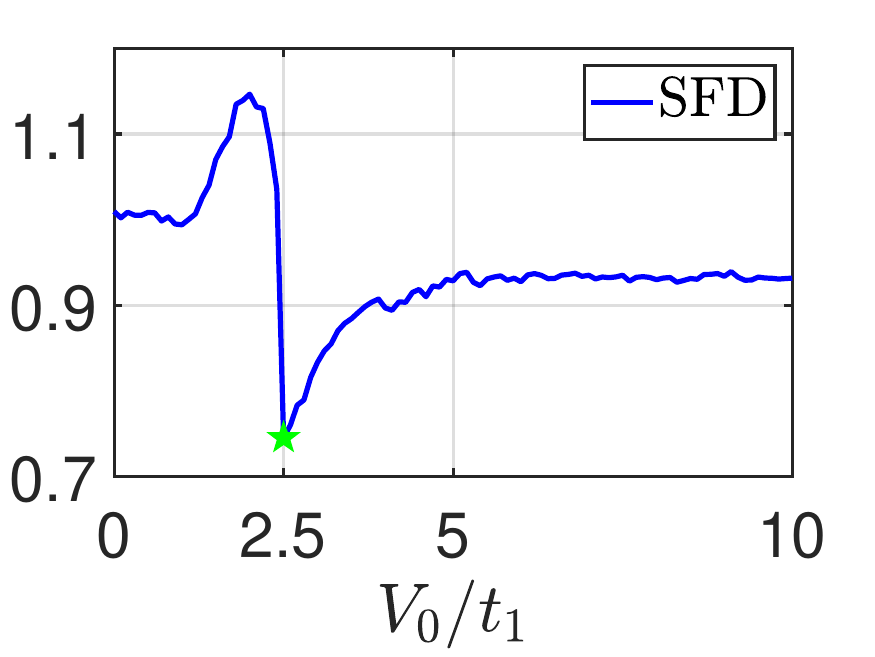}
    } \\
    \stackinset{l}{-0.1cm}{b}{2.55cm}{\colorbox{white}{(c)}}{
        \includegraphics[width=0.46\linewidth]{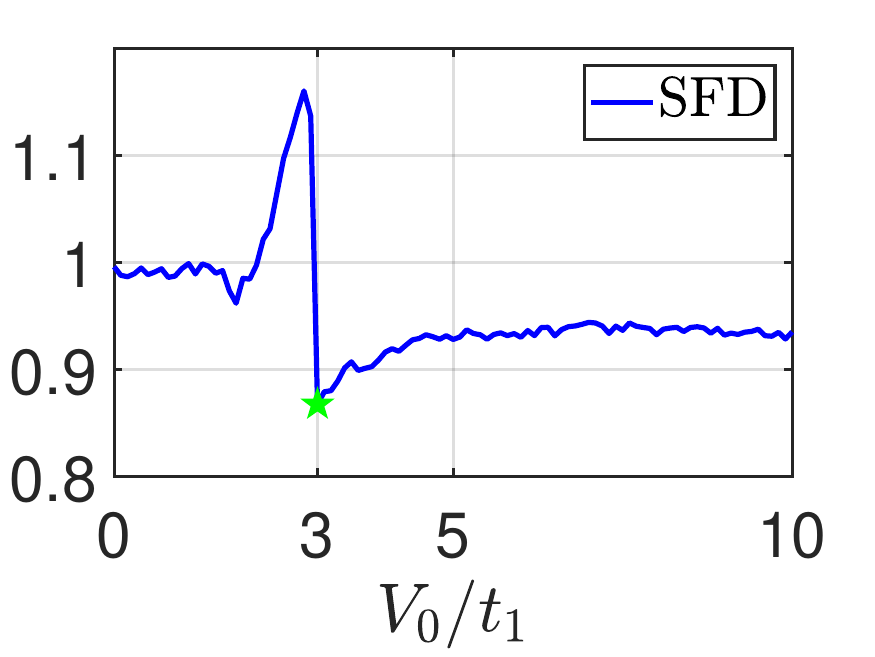}
    }
    \stackinset{l}{-0.1cm}{b}{2.55cm}{\colorbox{white}{(d)}}{
        \includegraphics[width=0.47\linewidth]{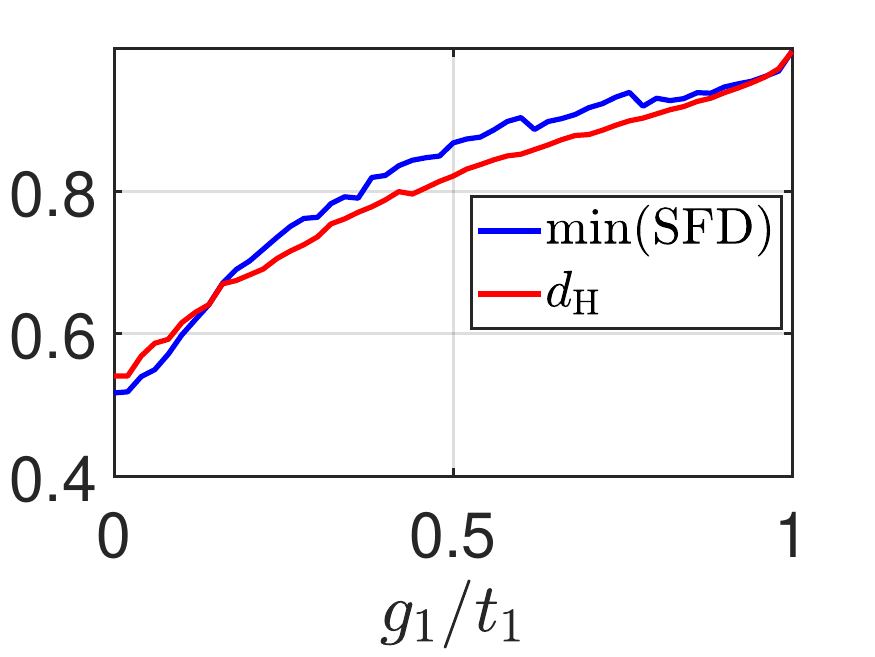} 
    }
    \caption{
    (a--c) SFD as a function of $V_0$ for $N=1597$.
    From Panels~(a) to (c), the parameters $(g_1, t_2, g_2 )$ are given by 
     $(0, 0, 0)$, 
    	$(0.25, 0, 0)$, 
    	 and   
    	$(0.5, 0, 0)$, respectively.
    The green stars mark the critical potential strengths $V_c /t_1 = 2$, $2.5$, and $ 3$ for Panels~(a)--(c), respectively, as deduced from  the averaged EFD in Fig.~\ref{Fig:EFD-vs-V0-g1}. 
    (d) Minimum of the SFD and fractal dimension $(d_{\rm H})$ of $p(E)$ as a function of $g_1$, where $V_0$ is chosen such that $V_0 = V_c (g_1)$ for the corresponding $g_1$ values.
      }
  \label{Fig:SFD-vs-V0_2}  
\end{figure}

\begin{figure}[t]
    \centering
    \stackinset{l}{-0.1cm}{b}{2.7cm}{\colorbox{white}{(a)}}{
        \includegraphics[width=0.485\linewidth]{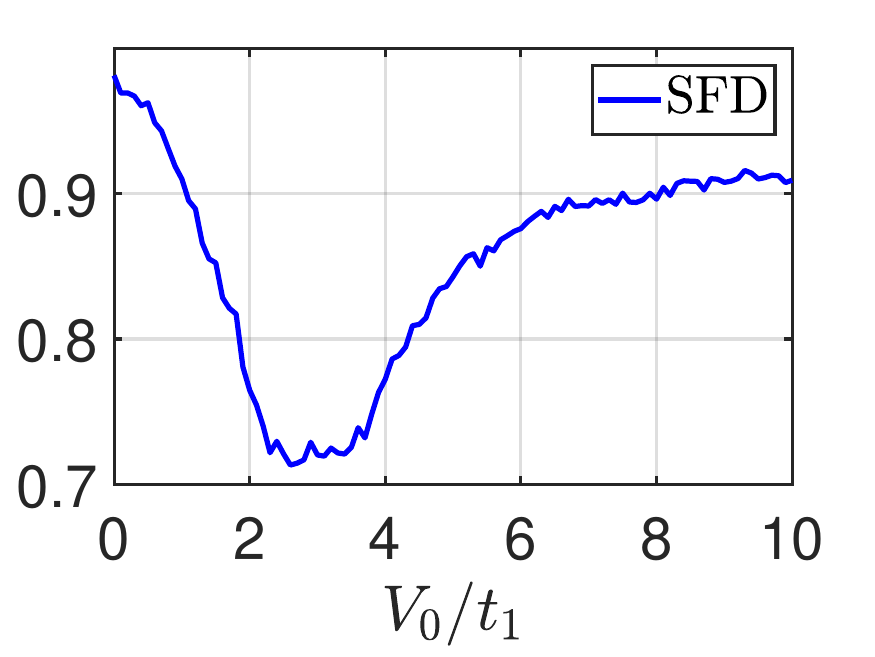}
    }
        \stackinset{l}{-0.1cm}{b}{2.7cm}{\colorbox{white}{(b)}}{
        \includegraphics[width=0.45\linewidth]{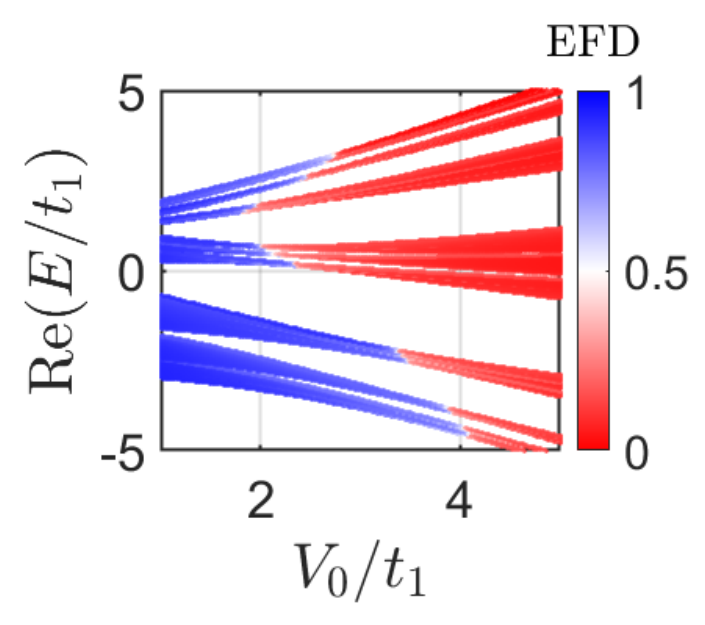}
    }
    \caption{(a) Similar to Fig.~\ref{Fig:SFD-vs-V0_2}(a)--(c) but for  $(g_1, t_2, g_2 ) = (0, 0.45, 0)$.
      (b) EFD ($\Gamma_n$) for the same parameter set as Panel~(a).
      }
  \label{Fig:SFD-vs-V0_3}  
\end{figure}

Its evolution with the potential strength $V_0$ for different sets of hopping parameters is presented in Fig.~\ref{Fig:SFD-vs-V0_2} and Fig.~\ref{Fig:SFD-vs-V0_3}(a).
In these plots, we find a unity SFD at $V_0 = 0$, corresponding to a spectrum without any onsite potential and forming either a closed loop (for $g_n \neq 0$) or a continuous solid line (for $g_n = 0$) on the complex plane. Both cases thus have the dimension of unity. 
Moreover, it is known that the fractal dimension of the \AAm model at the critical value $V_c = 2 t_1$ is roughly $0.5$~\cite{Tang:1986,Kohmoto:1987}.
It is noted that this value is reproduced in Fig.~\ref{Fig:SFD-vs-V0_2}(a),  with SFD $\sim 0.52$ at the critical value.
Additionally, we observe  bounded values of SFD as expected. Namely, in Fig.~\ref{Fig:SFD-vs-V0_2}(a) and Fig.~\ref{Fig:SFD-vs-V0_3}(a), the SFD is capped at unity because the system is in the Hermitian regime with a  real spectrum with a SFD between zero and unity. 
In contrast, in Fig.~\ref{Fig:SFD-vs-V0_2}(b,c), the SFD is not bounded by unity, since non-Hermiticity allows the spectrum to extend into the complex plane and to acquire a fractal dimension greater than one. More generally, for non-Hermitian lattices the spectrum resides in the two-dimensional complex plane; with quasiperiodicity-induced fractal structures, the associated fractal dimension can therefore take values between one and two.

Notably, one can find correlation between the SFD and $V_c$ deduced from the averaged EFD in Fig.~\ref{Fig:EFD-vs-V0-g1}(c,d). 
To be precise, in Fig.~\ref{Fig:SFD-vs-V0_2}(a)--(c), the SFD drops significantly toward the critical point and reaches its minimum at the respective $V_c$ (marked by green stars), indicating that the spectrum is the least continuous at the critical potential strength~\cite{Wu:2021,Theiler:1990}. 
This is true even for Fig.~\ref{Fig:SFD-vs-V0_2}(b,c), where a  nonzero $g_1$ is introduced, demonstrating the robustness of this behavior against non-Hermiticity.  
We also note that a larger $g_1$ value leads to a shift of the minima in Fig.~\ref{Fig:SFD-vs-V0_2}(a)--(c) toward a larger SFD value, consistent with the discussion in Ref.~\cite{Longhi:2021}.
In Fig.~\ref{Fig:SFD-vs-V0_2}(d), we further confirm this behavior for the entire range of $g_1/t_1 \in [0,1]$.

\begin{figure}[t]
    \centering
    \stackinset{l}{-0.1cm}{b}{2.55cm}{\colorbox{white}{(a)}}{
        \includegraphics[width=0.46\linewidth]{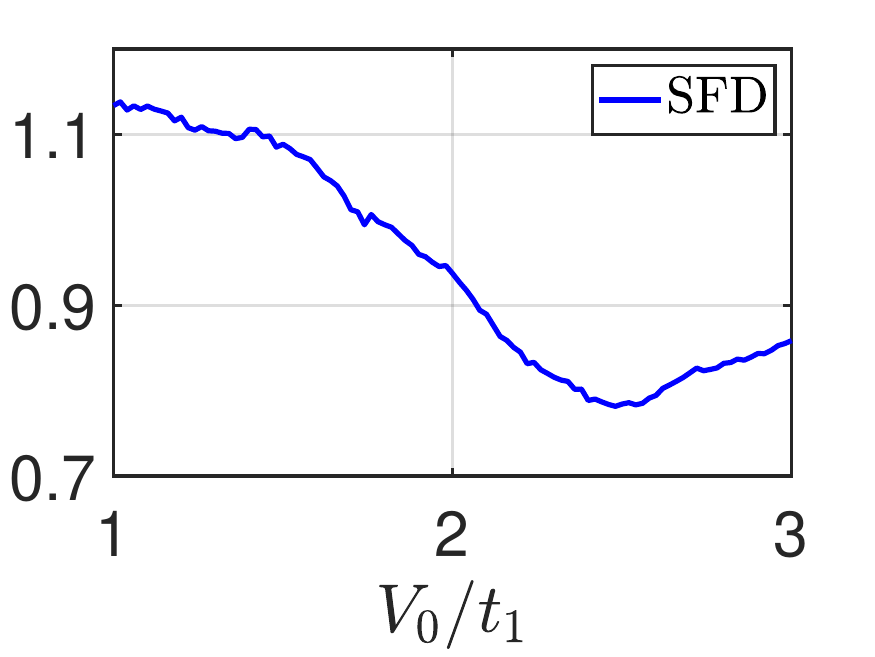}
    }
    \stackinset{l}{-0.1cm}{b}{2.55cm}{\colorbox{white}{(b)}}{
        \includegraphics[width=0.46\linewidth]{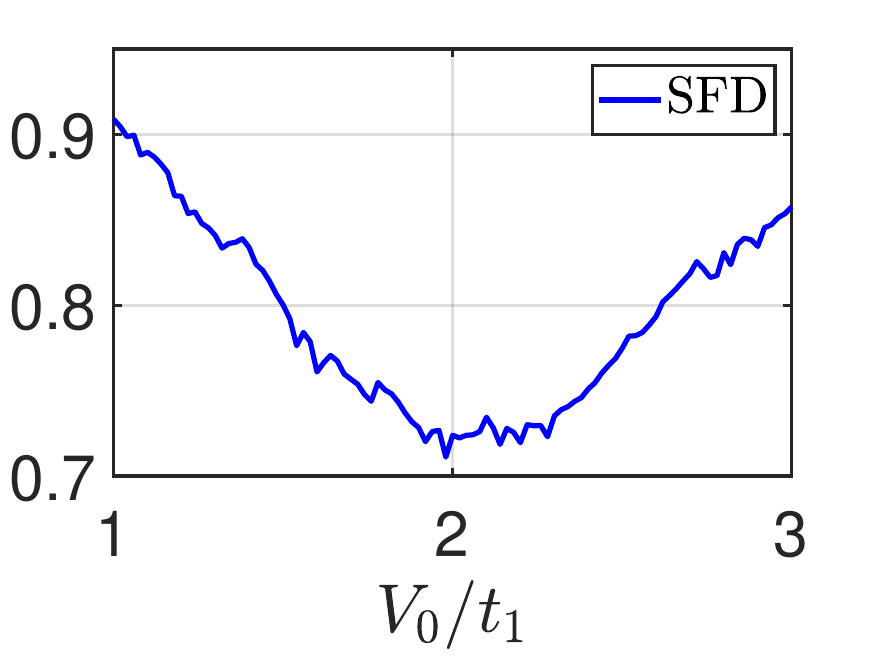}
    }
    \caption{SFD as a function of $V_0$ for $N=4181$.
    The other parameters $(g_1, t_2, g_2 )$  are given by 
      (a) $ (0.05, 0.1, 0)$ and (b) $(0, 0.1, 0)$.}
  \label{Fig:SFD-vs-V0_4}  
\end{figure}

\begin{figure*}[t]
    \centering 
    \stackinset{l}{-0.1cm}{b}{3.5cm}{\colorbox{white}{(a)}}{
        \includegraphics[width=0.315\linewidth]{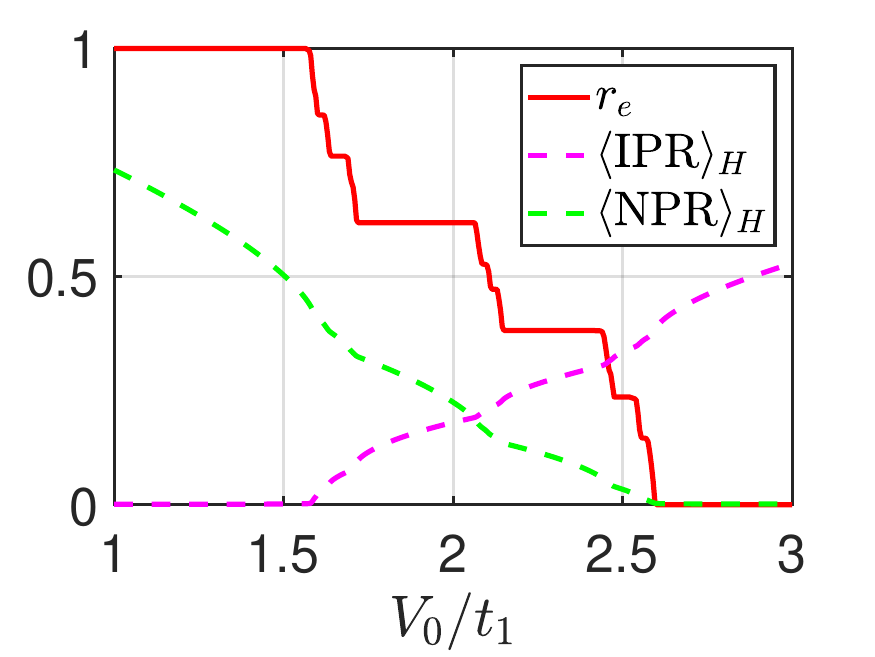}    
    }
    \stackinset{l}{-0.4cm}{b}{3.5cm}{\colorbox{white}{(b)}}{
        \includegraphics[width=0.315\linewidth]{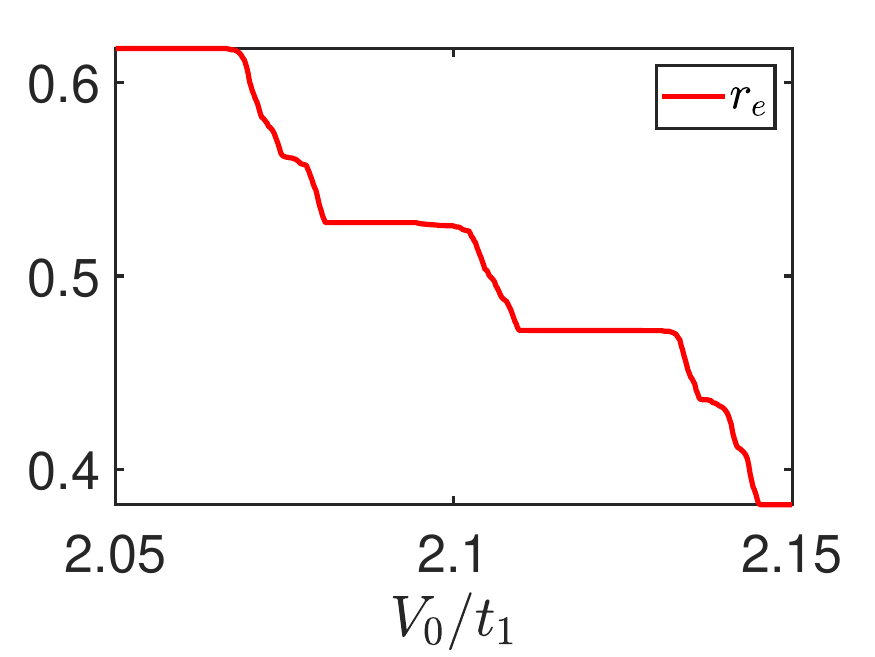}    
    }    
    \stackinset{l}{0cm}{b}{3.5cm}{\colorbox{white}{(c)}}{
        \includegraphics[width=0.315\linewidth]{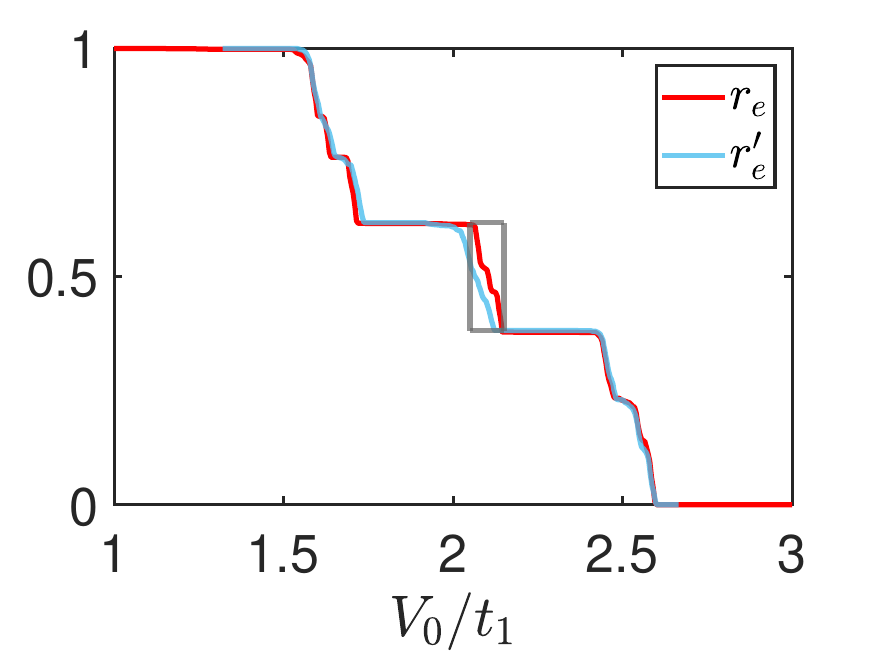}
    } 
    \caption{$\langle \text{IPR} \rangle_H$, $\langle \text{NPR} \rangle_H$, and EER ($r_e$) as functions of $V_0$, computed from  Eq.~\eqref{Eq:IPR}, Eq.~\eqref{Eq:NPR}, and Eq.~\eqref{Eq:EER}, respectively. 
    (a) Results for the resolution $\delta_{V_0} = 5 \times 10^{-3}$ and $N=1597$. 
    (b) Zoom-in view of $r_e$ for $ N=15000$ and $\delta_{V_0} = 5 \times 10^{-4}$ within $V_0 \in [2.05, 2.15]$. 
    (c) Comparison between $r_e$ and rescaled $r_e^\prime$, the latter being obtained through linear scaling and matching the leftmost and rightmost plateaus in the two curves.  
    The adopted values of the parameters are given by $(g_1,t_2,g_2) = (0.05,0.1,0)$. 
    } 
    \label{Fig:re_IPR_NPR} 
\end{figure*}

As shown in Fig.~\ref{Fig:EFD}, in the presence of longer-range hopping terms, a well-defined critical strength $V_c$ does not emerge to clearly separate the fully localized phase (where all eigenstates are localized) from the fully extended phase (where all eigenstates are extended). Notably, this feature can also be captured in SFD.
As shown in Fig.~\ref{Fig:SFD-vs-V0_3}(a), 
no sharp dip in the SFD is observed in this regime. 
Instead, after decreasing from $V_0/t_1 = 0$ to $V_0/t_1 = 2.3$, the SFD develops a flattened region over $V_0/t_1 \in [2.3,3.7]$ and then rises again for $V_0/t_1 > 3.7$. In Fig.~\ref{Fig:SFD-vs-V0_3}(b), we present the corresponding EFD plot, which reveals a mobility edge spanning the region where the SFD remains low and flat.

Before proceeding, we remark that the above correspondence between the flattened region in the SFD and the mobility edge applies only within a certain parameter regime. For demonstration, in Fig.~\ref{Fig:SFD-vs-V0_4}(a) we show the SFD using the same parameter set as in Fig.~\ref{Fig:EFD}. As expected for a system without a critical potential strength, no sharp dip appears; instead, a broadened, shallow minimum develops. Notably, due to non-Hermiticity, the SFD can exceed unity, and the resulting shallow minimum is less pronounced than the one in Fig.~\ref{Fig:SFD-vs-V0_3}(a). Consequently, a direct comparison between the SFD and the mobility edge extracted from Fig.~\ref{Fig:EFD} is not straightforward.
To enable a meaningful comparison, we compute the corresponding SFD in the Hermitian limit by setting $g_{1}=0$ and present the result in Fig.~\ref{Fig:SFD-vs-V0_4}(b). Again, no sharp dip is observed. However, a clear shallow, basin-like region appears in the range $V_{0}\in[1.9,2.2]$, which is located within the mobility-edge range in Fig.~\ref{Fig:EFD}. 

Interestingly, in addition to directly extracting fractal dimension from the complex energy spectra, we also uncover hidden self-similar features in our model, which we discuss in the following sections.

\subsection{Self-similar features in the extended eigenstate ratio (EER)}
\label{Sec:hidden}

In this section we further quantify the quasiperiodicity-induced transitions when the longer-range hopping terms are present.
Motivated by the observation in Fig.~\ref{Fig:PBC-Spectra_HNL-qp}, where extended eigenstates gradually vanish as the potential strength increases,
we introduce the EER, 
\begin{equation}
    r_e = \frac{N_e}{N}, 
    \label{Eq:EER}
\end{equation}
where $N_e$ is the number of eigenstates with the EFD, $\Gamma_n > 0.5$; see Eq.~\eqref{Eq:EFD} for the definition of EFD.
We use $r_e$ to estimate the fraction of (relatively) extended states present in the system. 
We remark on our choice of this criterion. In general, any $\Gamma_n $ value that does not converge to unity or zero, may correspond to fractal or power-law localized states. In our results [see, e.g., Fig.~\ref{Fig:EFD-vs-V0-g1} and Fig.~\ref{Fig:EFD}], clear mobility edges emerge, separating regions where EFD approaches unity from those where it approaches zero, and these boundaries tend to align with $\Gamma_n = 0.5$. This motivates our use of $\Gamma_n = 0.5$ as a practical numerical criterion. We emphasize, however, that this threshold should be regarded as a heuristic indicator rather than a universal definition, and in other systems or parameter regimes, it may not faithfully capture the precise number of extended states.
 
In Fig.~\ref{Fig:re_IPR_NPR}(a), we present the dependence of the EER on the potential strength, alongside the averaged IPR and NPR defined in Eq.~\eqref{Eq:IPR} and Eq.~\eqref{Eq:NPR}, respectively.
Consistent with Fig.~\ref{Fig:PBC-Spectra_HNL-qp}, where the extended states progressively vanish as $V_0$ increases, we observe  step-like feature in the $r_e$ curve. 
Additionally, since the $V_0$ value at which $\langle \text{IPR} \rangle_H = 0$ does not coincide with the value for $\langle \text{NPR} \rangle_H = 0$, there exists a finite range of $V_0$ within which localized and extended states coexist in the spectrum. 
Aligning with this, we identify two points in Fig.~\ref{Fig:re_IPR_NPR}(a): 
 $V_0 \approx 1.55$, below which $\langle \text{IPR} \rangle_H = 0$, and $V_0 \approx 2.6$, above which $\langle \text{NPR} \rangle_H = 0$, corresponding to the edges of the $r_e = 1$ and $r_e = 0$ plateaus, respectively.
Thus, by tracking the evolution of $r_e$ as a function of $V_0$, we gain insights into how the quasiperiodic potential drives the localization transition of the eigenstates.

Remarkably, the $r_e (V_0)$ curve provides another perspective on how quasiperiodicity influences the physical properties of the system.
Namely, Fig.~\ref{Fig:re_IPR_NPR} reveals self-similar structure in $r_e$ as $V_0$ varies. 
To explore this potential self similarity, we focus on the middle region of Fig.~\ref{Fig:re_IPR_NPR}(a), specifically for $V_0 \in [2.05, 2.15]$,  enhance the resolution $\delta_{V_0}$, and then increase the system size $N$ to obtain Fig.~\ref{Fig:re_IPR_NPR}(b). A comparison between Fig.~\ref{Fig:re_IPR_NPR}(a) and Fig.~\ref{Fig:re_IPR_NPR}(b) reveals a striking resemblance up to rescaling. This observation is further substantiated by rescaling $r_e (V_0)$ within $V_0 \in [2.05, 2.15]$ and overlaying the rescaled data, $r_e^{\prime}$, onto the original data, as shown in Fig.~\ref{Fig:re_IPR_NPR}(c). 
In addition to the self similarity, we remark that the EER curve also resembles the {\it Devil's staircase}~\cite{Robert:2021}. 
Namely, beyond 
the fact that this mapping continuously transforms a closed interval into another closed interval, we also find that the curve is monotonically decreasing and remains constant almost everywhere.
Following the algorithm in Ref.~\cite{Jensen:1983}, we project the $r_e(V_0)$ curve in Fig.~\ref{Fig:re_IPR_NPR}(a) onto the $V_0$ axis, interpreting the plateaus as ``empty boxes'' and the sloped regions as ``filled boxes.''
After this projection, we apply the box-counting method to extract the associated fractal dimension, $d_{\rm H}  \approx 0.56$. 
As a self-similar feature, the positions of the two widest plateaus is related to the quasiperiodicity set by $\lambda_0/a_0$, which will be discussed below.
We remark that the self-similar structure in the $r_e$ curve emerges only when longer-range hopping terms are included. 
In their absence, a critical potential strength exists and the EER exhibits only a single discontinuity at the critical potential strength, with no indication of additional plateaus.
Furthermore, since introducing the non-Hermitian term results in complex energy spectrum, it can induce  self-similar structures in the complex eigenvalue ratio introduced in Ref.~\cite{Padhi:2024}; see its definition in Eq.~\eqref{Eq:CER} and related discussion in Appendix~\ref{Appx:r_c}.

To gain deeper insight into the origin of this self similarity, we now revisit the system at criticality and explore how self-similar features can arise in alternative spectral quantities, and how they relate to the structures observed in the presence of longer-range hopping discussed here.

\subsection{Self similarity close to the criticality }
\label{Sec:ss}

To gain insight into the origin of the self similarity, we begin by examining the system without longer-range hopping, where a well-defined criticality exists. At this critical point, the energy spectrum remains real~\cite{Longhi:2021}, allowing us to define the spectral survival ratio as
\begin{equation}
    p(E) = n(E)/N,
    \label{Eq:percent_eig}
\end{equation}
where $n(E)$ denotes the number of eigenvalues greater than the threshold energy $E$, and $N$ is the total number of eigenvalues.
The spectral survival ratio thus quantifies the fraction of eigenstates above an energy threshold $E$. Interestingly, for a given spectrum, the $p(E)$ curve can be interpreted as casting the spectrum into a two-dimensional structure in the $p$-$E$ space. A notable example is provided by the Aubry-Andr{\'e} criticality at $g_1 = 0$, where the energy spectrum resembles a Cantor set~\cite{Hofstadter:1976}, and the corresponding $p(E)$ curves naturally exhibit self-similar features; see Appendix~\ref{Appx:fractal} for a brief review. 

\begin{figure}[t]
    \centering
    \stackinset{l}{-0.4cm}{b}{2.4cm}{\colorbox{white}{(a)}}{
        \includegraphics[width=0.465\linewidth]{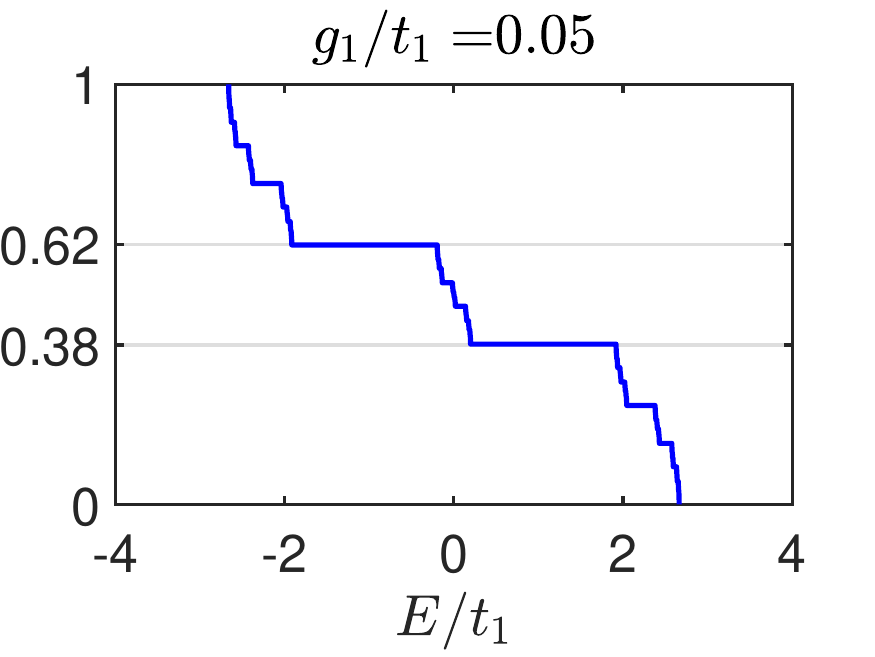}
    }
    \stackinset{l}{-0.4cm}{b}{2.4cm}{\colorbox{white}{(b)}}{
        \includegraphics[width=0.465\linewidth]{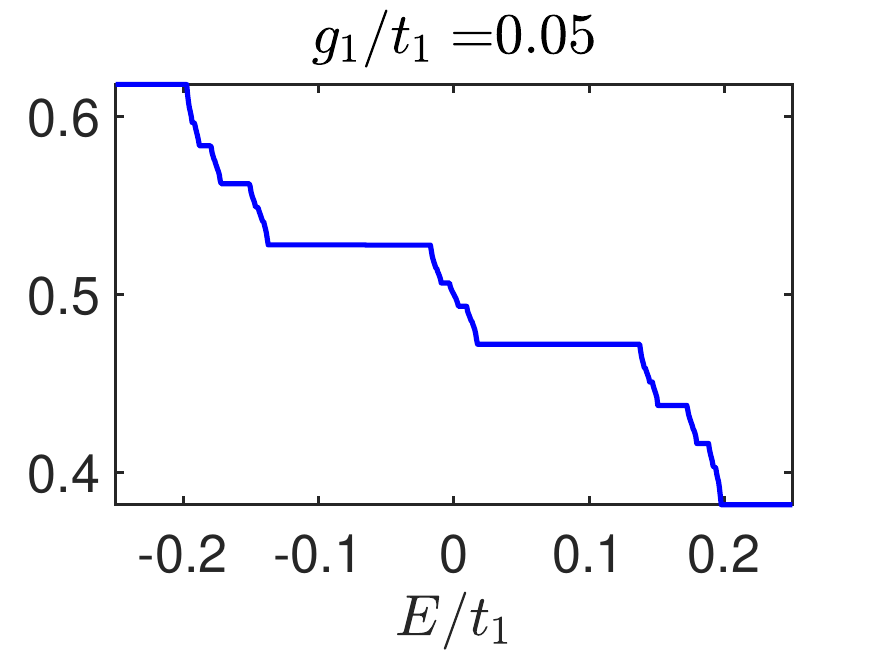} 
    }
    \caption{ (a) Spectral survival ratio $p(E)$ of the critical spectrum for $N=1597$, $(g_1, t_2, g_2, V_0) = (0.05,0,0, 2.1)$. (b) Zoom-in view of Panel~(a) using $N=8000$.
    }
    \label{Fig:p(E) and zoom_in}
\end{figure}

With a nonzero $g_1$ value, we go beyond the Aubry-Andr{\'e} criticality but can still obtain a critical spectrum by setting $V_0 = V_c(g_1)$. The spectrum remains real under this condition, and the corresponding spectral survival ratio is shown in Fig.~\ref{Fig:p(E) and zoom_in}. Although the critical spectrum no longer resembles a Cantor set and the sharp features in the $p(E)$ curves are slightly smeared out (compared to Fig.~\ref{Fig:p(E) and zoom_in-Appx} in Appendix~\ref{Appx:fractal}), it still exhibits self-repeating plateau structures. Since a nonzero $g_1$ tends to smooth out fine features in $p(E)$, a larger system size $N$ is required to resolve the underlying self-similar structure; see Fig.~\ref{Fig:p(E) and zoom_in}(b).
 
Remarkably, we find that the fractal dimension $d_{\rm H}$ of $p(E)$ remains systematically consistent\footnote{The minor discrepancies arise from the procedure of ``flattening'' the $p(E)$ curves, as well as differences between the method used to extract $d_{\rm H}$ from the resulting patterns~\cite{Jensen:1983} and the box-counting approach employed for the SFD~\cite{Theiler:1990}.}
with the minimum of the SFD extracted from the PBC spectra over a range of $g_1$, 
as shown in Fig.~\ref{Fig:SFD-vs-V0_2}(d). 
Furthermore, as $g_1$ increases, the critical spectrum becomes progressively more continuous, eventually reaching an SFD of unity at $g_1 = t_1$; see Appendix~\ref{Appx:fractal} for a more detailed discussion.

Before concluding this section, we note that one can define another related yet distinct quantity--the complex eigenvalue ratio, denoted by $r_c$~\cite{Padhi:2024}; see Appendix~\ref{Appx:r_c}. As shown in Fig.~\ref{Fig:rc_IPR_NPR}, $r_c$ also exhibits self-similar features, and its extracted fractal dimension is approximately $d_{\rm H} \approx 0.57$, closely matching that of $r_e$. We further find that the spectral survival ratio $p(E)$ at $g_1 = 0.05$ gives a similar fractal dimension; see Fig.~\ref{Fig:SFD-vs-V0_3}(b).

Interestingly, the positions of the two widest plateaus in $r_e$, $p(E)$, and $r_c$ are closely related to the irrational periodicity $\lambda_0/a_0$.
Following Ref.~\cite{Hofstadter:1976} and the discussions in Appendix~\ref{Appx:fractal}, we closely examine Fig.~\ref{Fig:p(E) and zoom_in}, where $g_1 \neq 0$ and $V_0 = V_c(g_1)$, such that $p(E)$ remains well-defined.
We find that for a local variable $\beta$, defined as the fractional part of $\lambda_0/a_0$ [that is, with $\beta = (\sqrt{5} - 1)/2 \approx 0.62$ in this case], the two widest plateaus occur at $p(E) = \beta$ and $p(E) = 1 - \beta$.
This behavior is expected to persist when longer-range hopping terms are finite but sufficiently small, as is indeed observed in Fig.~\ref{Fig:re_IPR_NPR}.

In summary, our results uncover self-similar structures reminiscent of the Devil's staircase across multiple localization-related quantities, providing insights into the nature of localization in non-Hermitian systems.

\section{Discussion}
\label{Sec:discussion}

Our work reveals intriguing topological and transport properties in non-Hermitian quasiperiodic systems, which go beyond the well known skin effect. 
In the absence of longer-range hopping, the critical strength of the onsite potential exhibits a linear dependence on the antisymmetric hopping amplitude [see Fig.~\ref{Fig:EFD-vs-V0-g1}(d)]. In this regime, we also identify a dip in SFD at the critical potential strength [see Fig.~\ref{Fig:SFD-vs-V0_2}(a)--(c)], 
establishing a direct relation between this spectral measure and the critical strength.
The existence of the critical potential strength allows us to define the spectral survival ratio, which reveals self similarity. 

Our analysis further contributes to the understanding of generalized mobility edges in non-Hermitian systems. Existing studies have examined models with Hermitian hopping and a complex potential, revealing the emergence of mobility edges in the real part of energy spectra~\cite{Liu:2020a,Xia:2022}  
or deriving an analytical expression for the mobility ring in the complex spectrum~\cite{Li:2024}. Additionally, Ref.~\cite{Zeng:2020a} investigated a system with both complex asymmetric hopping and a complex potential, identifying the mobility edges numerically.  
Here, our analysis demonstrates that the concept of mobility edges can be effectively applied in non-Hermitian systems by characterizing them using the real part of the PBC spectrum. Aligning with Ref.~\cite{Zeng:2020a}, our numerical results suggest that as long as localized and extended states can be clearly distinguished, the definition of the mobility edge may not be unique.

The inclusion of longer-range hopping not only enriches the topological properties but also gives rise to novel localization phenomena. The introduction of $t_2$ and $g_2$ has several effects. In the absence of an onsite potential, it leads to higher winding numbers; see Figs.~\ref{Fig:nh}--\ref{Fig:H_HNL_Spectra}. When an onsite potential is present, it suppresses the sharp drop in SFD as a function of potential strength [see Fig.~\ref{Fig:SFD-vs-V0_3}(a)] 
and induces the emergence of a mobility edge (see Fig.~\ref{Fig:EFD}).
These terms also lead to emergence of self-similar features in various quantities related to the quasiperiodicity-induced localization.

The self-similar structures observed in  the extended  eigenstate ratio (see Fig.~\ref{Fig:re_IPR_NPR}), the spectral survival ratio (see Fig.~\ref{Fig:p(E) and zoom_in}), and the complex eigenvalue ratio (see Fig.~\ref{Fig:rc_IPR_NPR}) 
suggest that fractal or self-similar features may be a general property of non-Hermitian quasiperiodic systems, as seen in other settings~\cite{Sun:2024}.
While Ref.~\cite{Padhi:2024} computed quantities similar to those presented here, and some Hermitian settings have shown patterns that hint at self-similar features~\cite{Biddle:2010,Peng:2024}, 
none of these studies explored the possibility of hidden self-similar features. 
This omission may be attributed to the relatively small system sizes and low resolution (limited meshing points) used in their numerical analyses.  
Our findings go beyond the typical fractal structures observed in the spectrum and density of states for systems with fixed parameters. Instead, we reveal that fractal-like features can also emerge when varying system parameters, drawing a parallel to the Hofstadter butterfly~\cite{Hofstadter:1976}, where fractality arises as a function of the magnetic flux.
 
Several aspects warrant further exploration. The spectral winding number of the PBC spectrum appears to be correlated with the localization properties of the eigenstates. While our numerical analysis reveals a linear relationship between $V_c$ and $g_1$, we suspect that an analytic derivation may be feasible when longer-range hoppings are absent;
for the special case $|t_1/ g_1|=1$, in Ref.~\cite{Longhi:2021} the relation between $V_c$ and $g_1$ was demonstrated by analyzing the Lyapunov exponent, which also shows a linear $V_c$-to-$g_1$ relation.
Extending the analytic derivation to arbitrary $|t_1/ g_1|$
could offer deeper theoretical insight into the nature of localization transitions in non-Hermitian quasiperiodic systems. 

Finally, we note that our model can be feasibly realized in experimental platforms characterized by nonreciprocal elements, including ring resonators~\cite{Wang:2021}, electric circuits~\cite{Shao:2021,Helbig:2020}, and ultracold atoms with laser beams~\cite{Modugno:2010}. 
In addition, the nonreciprocal hoppings might equivalently describe a macroscopic array of asymmetric loops formed by conducting molecules, as recently explored in the context of non-unitary quantum devices in Ref.~\cite{Mannhart:2021}.
Additionally, Ref.~\cite{Padhi:2024} proposed an electric circuit setup that could be closely related to our system, offering a promising avenue for future experimental verification.

\section*{Data availability statement}

The data that support the findings of this article are openly available~\cite{data}.

\begin{acknowledgments}
 
We thank H.-C.~Kao, Y.~Kato, C.-K. Shih, and H.-C. Wang for interesting discussions.
We acknowledge support from the National Science and Technology Council (NSTC), Taiwan through Grant No.~NSTC-112-2112-M-001-025-MY3 and Grant No.~NSTC-114-2112-M-001-057, and Academia Sinica (AS), Taiwan through Grant No.~AS-iMATE-114-12.
We acknowledge the technical support from AS Grid Computing Center (ASGC), Taiwan through Grant No.~AS-CFII-112-103.
R.O. is supported by JSPS KAKENHI Grants No.~23K13033 and No.~24K00586.
Y.~W. acknowledges support from the Summer Student Research Program of the Institute of Physics, AS, Taiwan. 
C.-H.H. acknowledges support from the National Center for Theoretical Sciences (NCTS), Taiwan.

\end{acknowledgments}

\appendix

\section{Spectral and topological properties in the absence of the onsite potential }
\label{Appx:Spectral} 
 
In this section, we revisit the generalized Hatano-Nelson model in Eq.~\eqref{Eq:H_nH+qp} with  $V_0=0$ and $n_{\rm max} = 2$, and discuss their spectral winding number and skin effects for completeness. 
Performing Fourier transform, we obtain the PBC energy spectrum,  
\begin{eqnarray}
    E_{\rm nH}^{\PBC} (k) &=& - \big[ (t_1+g_1)e^{-ik a_0} + (t_1-g_1)e^{ik a_0} \nonumber  \\
   && + (t_2+g_2)e^{-2ik a_0} + (t_2-g_2)e^{2ik a_0} \big] ,
\label{Eq:E_HNL_PBC}    
\end{eqnarray}
where the last two terms arise from the longer-range hopping terms.

 \begin{figure}[t]
    \centering
    \stackinset{l}{-4pt}{b}{2.6cm}{\colorbox{white}{(a)}}
    {\includegraphics[width=0.493\linewidth]{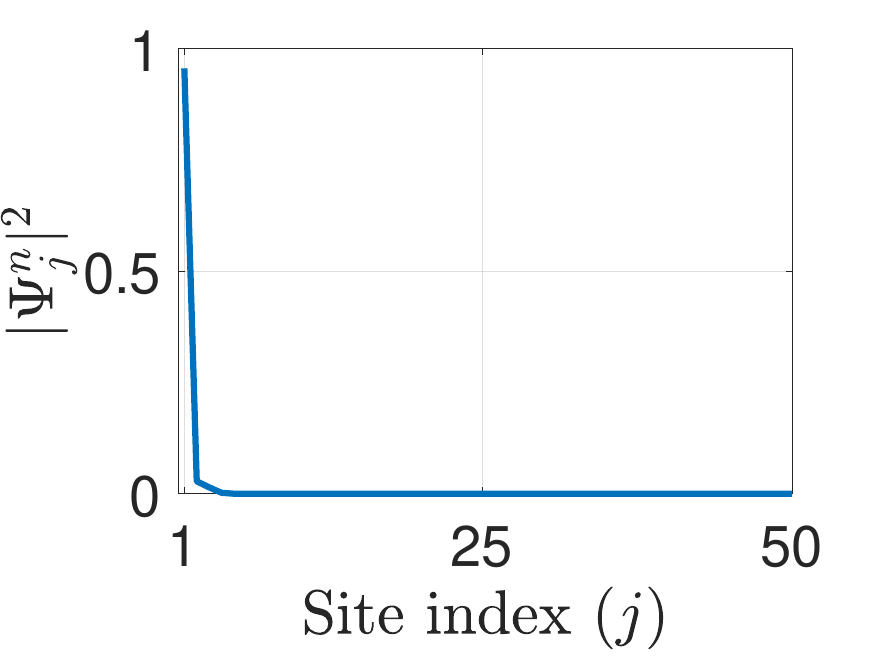}}
    \stackinset{l}{-6.5pt}{b}{2.6cm}{\colorbox{white}{(b)}}
    {\includegraphics[width=0.493\linewidth]{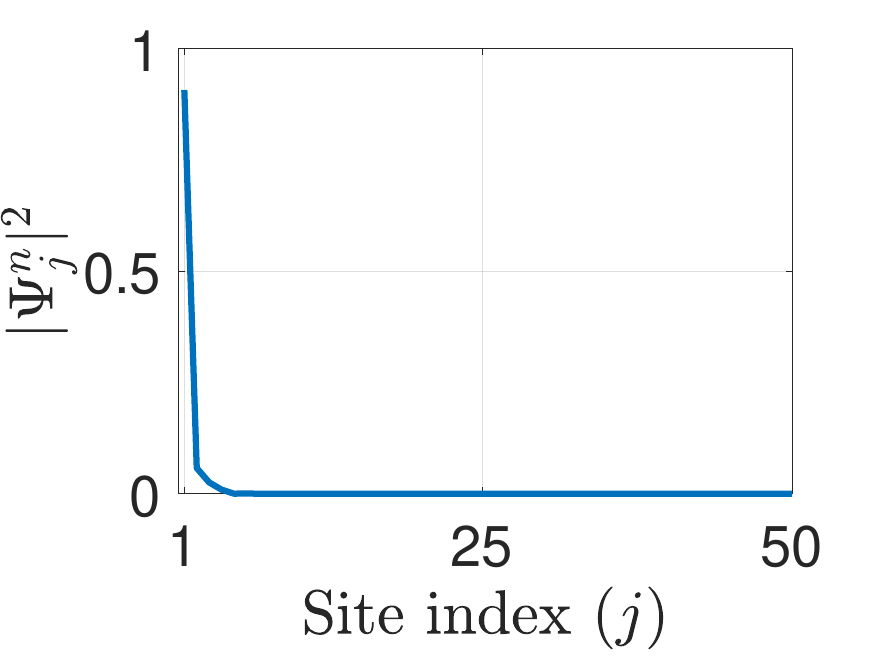}}
    \caption{(a,b) Spatial profiles for the eigenstates of $n = 22$ and $n = 50$ corresponding to Fig.~\ref{Fig:nh}, with their corresponding eigenvalues located inside the $W=1$ and $W=2$ regions, respectively.  
    Due to the antisymmetric hopping, these eigenstates tend to localize at one of the open boundaries (here, the left end).
    The adopted parameter values are the same as those in Fig.~\ref{Fig:nh}. 
    }
    \label{Fig:nh2}
\end{figure}

Under PBC, the energy spectrum forms closed loops in the complex energy plane from which one can define the winding number as in Eq.~\eqref{Eq:Winding}. 
As presented in the main text, Fig.~\ref{Fig:nh}(a) and Fig.~\ref{Fig:H_HNL_Spectra} illustrate the emergence of higher winding number $|W| > 1$. 
As expected, the longer-range hoppings make the energy spectrum $E_{\rm nH}^{\PBC}$ more complicated than ellipses.
Upon introducing the onsite potential, the PBC loops become distorted and eventually fragment into multiple loops. Under stronger potential strength, as shown in Fig.~\ref{Fig:PBC-Spectra_HNL-qp} in the main text, these loops progressively shrink and collapse onto the real axis. These behaviors are quantified using the EER or complex eigenvalue ratio, as discussed in the main text and the following section.

As reviewed in the main text, a distinctive characteristic of non-Hermitian systems is that the appearance of skin modes under OBC\footnote{For numerical calculations involving non-Hermitian matrices under OBC, the eigenvalue problem tends to be ill-conditioned, making it highly sensitive to perturbations. Consequently, larger system sizes are more prone to numerical errors~\cite{Reichel:1992}. 
As a result, a relatively small system size is typically chosen for computations under OBC.}, as shown in Fig.~\ref{Fig:nh2}. 
To examine the unique features arising from quasiperiodicity-induced bulk localization (i.e., due to $V_0 \neq 0$), we impose PBC in our numerical analysis throughout the main text, where such skin modes are suppressed.

\section{Complex eigenvalue ratio 
\label{Appx:r_c} }  

\begin{figure}[t]
    \centering
    \stackinset{l}{-0.2cm}{b}{2.5cm}{\colorbox{white}{(a)}}{
        \includegraphics[width=0.465\linewidth]{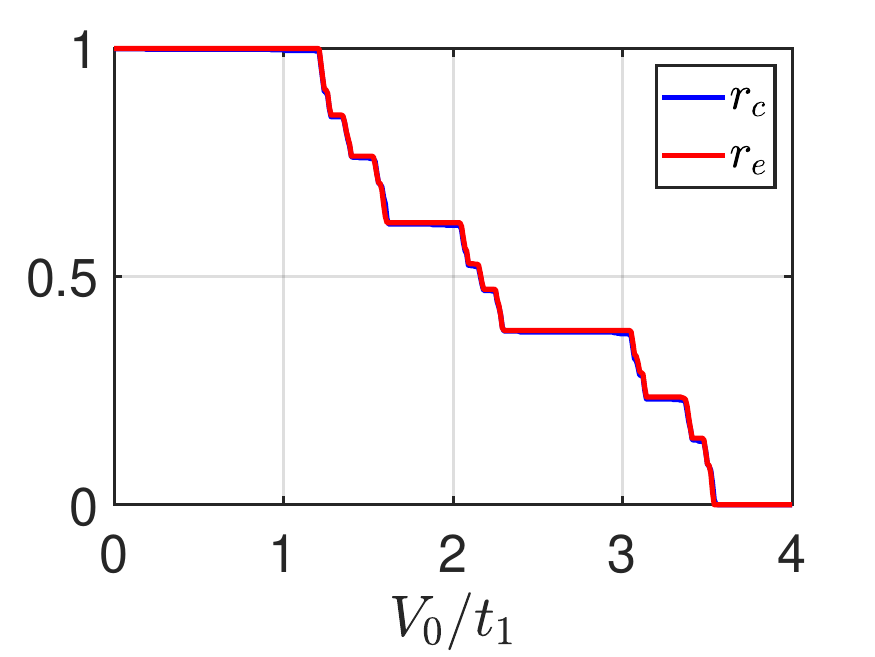}
    }
    \stackinset{l}{-0.2cm}{b}{2.5cm}{\colorbox{white}{(b)}}{
        \includegraphics[width=0.465\linewidth]{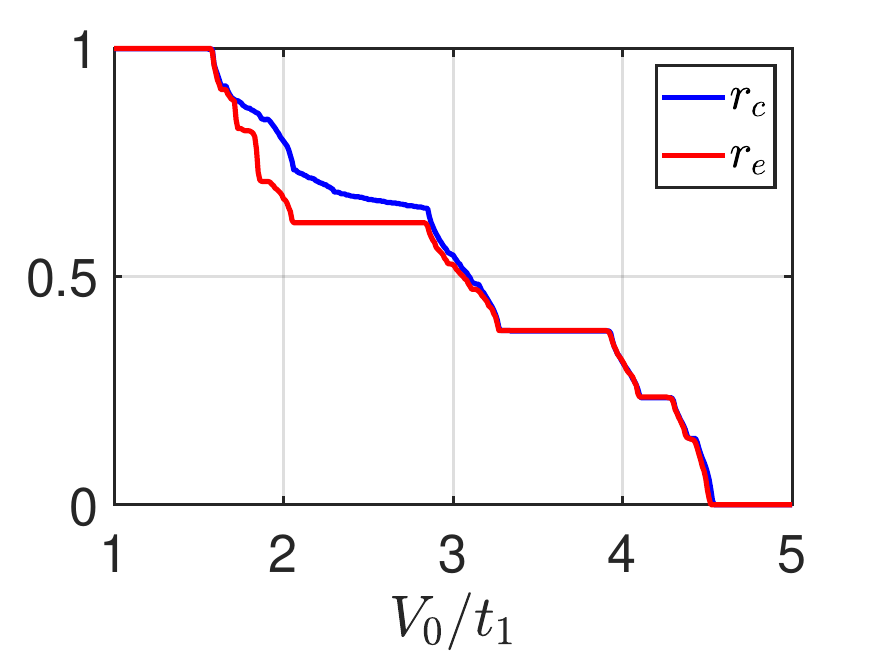} 
    }
    \caption{Comparison of $r_c$ and $r_e$ for $N=1597$, $\delta_{V_0} = 10^{-2}$; $r_c$ is computed from   Eq.~\eqref{Eq:CER}. For Panels~(a) and (b), the parameters $(g_1, t_2, g_2)$ are given by 
    $(0.05, 0.3,0)$  and $(0.5,-0.25,-0.1)$, respectively. 
    }
    \label{Fig:r_e r_c}
\end{figure}

In this section we demonstrate that self-similar structures can also arise from the complex eigenvalue ratio~\cite{Padhi:2024}, 
\begin{equation}
    r_c = \frac{N_c}{N},
    \label{Eq:CER}
\end{equation}
where $N_c$ is the number of eigenstates with nonzero imaginary part of eigenvalues. 
To understand how this quantity is related to the localization properties, we observe Fig.~\ref{Fig:PBC-Spectra_HNL-qp}(a)--(c), where eigenstates with complex eigenvalues are mostly extended and those with real eigenvalues are mostly localized.  To better demonstrate their differences, we plot $r_c$ and $r_e$ together in Fig.~\ref{Fig:r_e r_c}. In the absence of the longer-range antisymmetric hopping (that is, $g_2=0$), all the localized states have real eigenvalues, and $r_c$ and $r_e$ are essentially identical; see Fig.~\ref{Fig:r_e r_c}(a).  
On the other hand, for $g_2 \neq 0$, we obtain  distinct $r_c$ and $r_e$ curves, with the latter better capturing the staircase-like structure; see Fig.~\ref{Fig:r_e r_c}(b). In consequence, the EER works better in more general cases. In the main text we thus introduce the EER, $r_e$, rather than $r_c$, in the characterization, in order to cover broader parameter space where not all the eigenvalues localized states are real.

\begin{figure}[b]
    \centering 
    \stackinset{l}{-0.1cm}{b}{3.65cm}{\colorbox{white}{(a)}}{
        \includegraphics[width=0.65\linewidth]{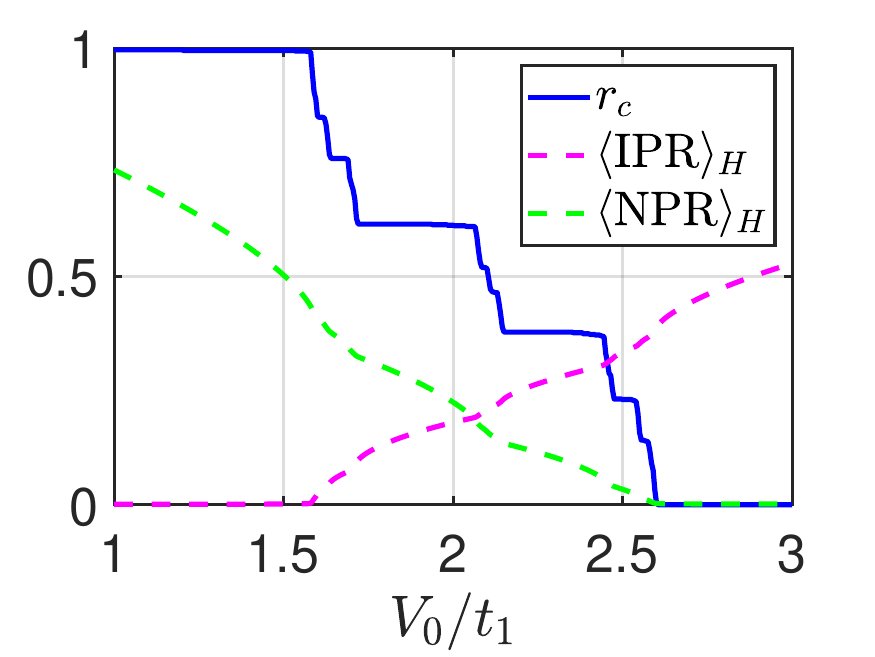}    
    } \\ 
    \stackinset{l}{-0.4cm}{b}{2.25cm}{\colorbox{white}{(b)}}{
        \includegraphics[width=0.45\linewidth]{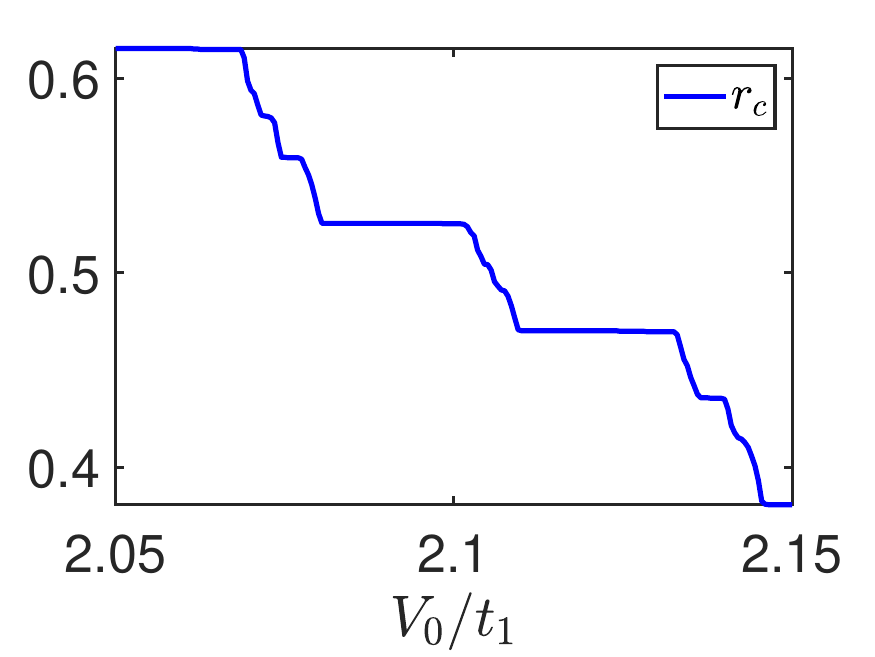}    
    }    
    \stackinset{l}{-0.2cm}{b}{2.25cm}{\colorbox{white}{(c)}}{
        \includegraphics[width=0.45\linewidth]{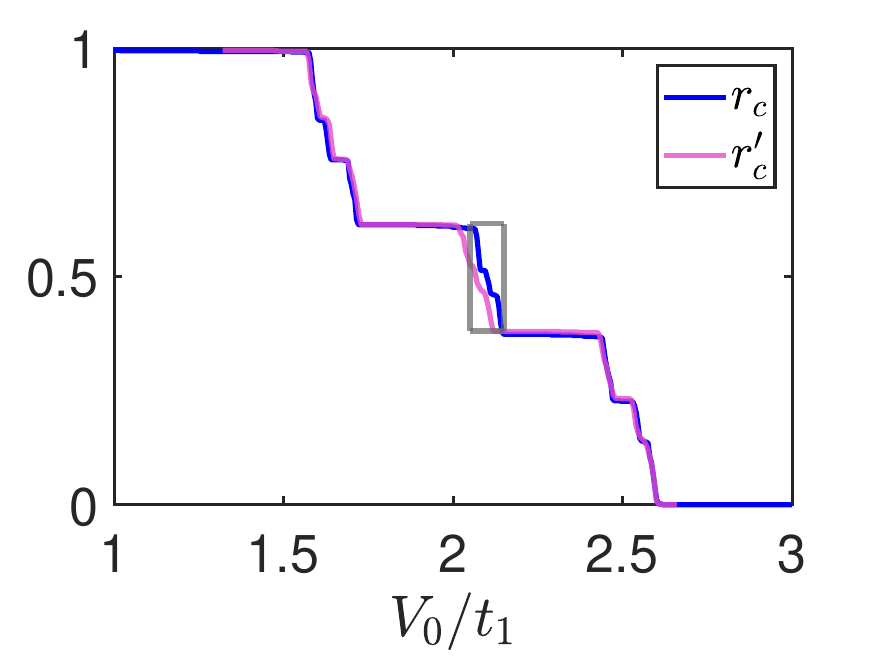}
    } 
    \caption{$\langle \text{IPR} \rangle_H$, $\langle \text{NPR} \rangle_H$, and $r_c$ as functions of $V_0$ for $N=1597$.
    (a) Results for the resolution $\delta_{V_0} = 5 \times 10^{-3}$. 
    (b) Zoom-in view of $r_c$ for $N=8000, \delta_{V_0} = 5 \times 10^{-4}$ within $V_0 \in [2.05, 2.15]$. 
    (c) Comparison between $r_c$ and rescaled $r_c^\prime$, obtained through linear scaling.  
    The adopted values of the parameters are given by $(g_1,t_2,g_2) = (0.05,0.1,0)$.
    }
    \label{Fig:rc_IPR_NPR} 
\end{figure}

\begin{figure}[t]
    \centering
    \stackinset{l}{-0.05cm}{b}{2.4cm}{\colorbox{white}{(a)}}{
        \includegraphics[width=0.46\linewidth]{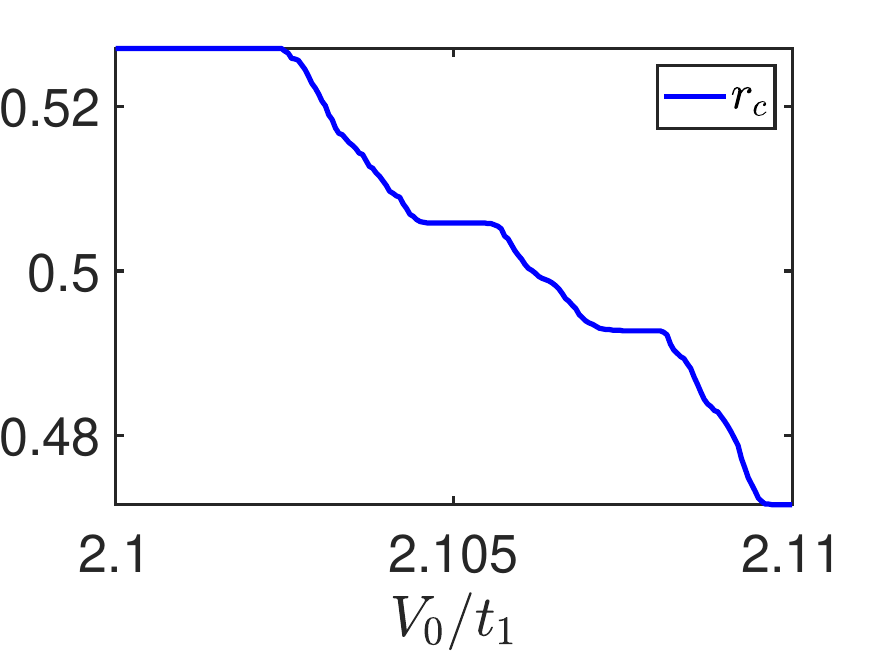}
    }
    \stackinset{l}{-0.05cm}{b}{2.4cm}{\colorbox{white}{(b)}}{
        \includegraphics[width=0.46\linewidth]{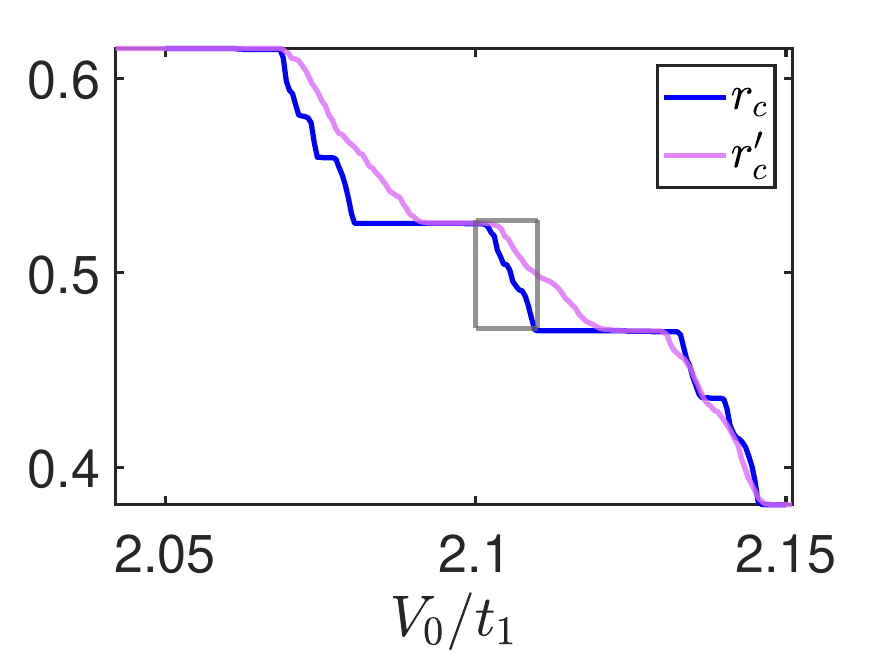}
    }
    \caption{Further zoom-in view of $r_c$ in analogous to Fig.~\ref{Fig:rc_IPR_NPR}.
    (a) Results for $\delta_{V_0}=5 \times 10^{-5}$, $N=25000$, and $(g_1,t_2,g_2) = (0.05, 0.1, 0)$. (b) Comparison between $r_c$ and $r'_c$, similar to Fig.~\ref{Fig:rc_IPR_NPR}(c) but with a smaller scale. 
   }
    \label{Fig:rc_closer_view}
\end{figure}

We now demonstrate that, in the applicable regime, $r_c$ also exhibits self-similar structures.  
In Fig.~\ref{Fig:rc_IPR_NPR}(a), we present the dependence of the complex eigenvalue ratio, $r_c$, on the potential strength, $V_0$, along with the averaged IPR and NPR. Resembling $r_e$ in the main text, the profile of $r_c$ also exhibits step-like features, originating from the collapse of PBC loops with increasing $V_0$. As done for $r_e$ in the main text, we zoom into the region $V_0 \in [2.05, 2.15]$ in Fig.~\ref{Fig:rc_IPR_NPR}(a) and enhance the resolution $\delta_{V_0}$ to obtain Fig.~\ref{Fig:rc_IPR_NPR}(b). A direct comparison between Fig.~\ref{Fig:rc_IPR_NPR}(a)--(b) again reveals a self-similar structure, as shown in Fig.~\ref{Fig:rc_IPR_NPR}(c), along with features resembling a Devil's staircase.
Figure~\ref{Fig:rc_closer_view} presents a further zoom-in of $r_c$, which again exhibits the three main plateaus similar to those observed prior to rescaling.
The slight deviations from perfect self-similarity at this scale are likely due to limited resolution in the energy spectrum; an even larger system size $N$ may be required to resolve finer features numerically.

\section{Detailed analysis on the self-similar structures} 
\label{Appx:fractal}

In this section, we carry out a more detailed analysis of $r_e$ and $r_c$, as presented in Figs.~\ref{Fig:re_IPR_NPR} and \ref{Fig:rc_IPR_NPR}, to further elucidate the origin of the observed self-similar structures. We separately examine systems with only nearest-neighbor hopping and those with additional longer-range terms, corresponding to the presence and absence of criticality, respectively.

\subsection{Systems with only nearest-neighbor hopping terms}

As a starting point, when only $t_1$ is nonzero, our model at the critical point $V_0 = V_c = 2t_1$ reduces to the well-known cases previously studied by Hofstadter~\cite{Hofstadter:1976} and by Aubry and Andr{\'e}~\cite{Aubry:1980}. These systems are known to exhibit fractal structures in their energy spectra, such as the Hofstadter butterfly and the critical spectrum of the Aubry-Andr{\'e} model. In this setting, the spectrum can be characterized by a local variable $\beta$, determined by the fractional part of $\lambda_0/a_0$ in our model.
As discussed in Ref.~\cite{Hofstadter:1976}, each parent spectrum splits into three child spectra--left (L), right (R), and central (C)--each related to the parent through its local variable. Denoting the local variables of the parent spectrum, the L and R child spectra, and the C child spectrum by $\beta$, $\alpha'$, and $\beta'$, respectively, one finds that when $\lambda_0/a_0 = (\sqrt{5} + 1)/2$ and thus $\beta = (\sqrt{5} - 1)/2$, the parent and all three child spectra share the same local variable $\beta$. This implies that the band-gap structures are arranged identically across generations and can be linearly mapped onto one another, thereby exhibiting self similarity.

\begin{figure}[t]
    \centering
    \stackinset{l}{-0.2cm}{b}{2.6cm}{\colorbox{white}{(a)}}{
        \includegraphics[width=0.465\linewidth]{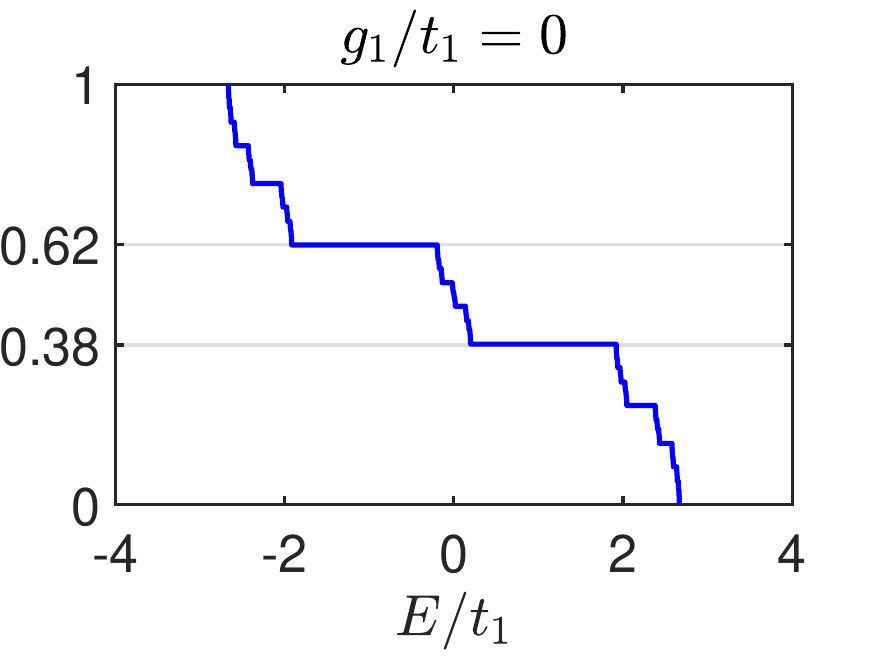}
    }
    \stackinset{l}{-0.2cm}{b}{2.6cm}{\colorbox{white}{(b)}}{
        \includegraphics[width=0.465\linewidth]{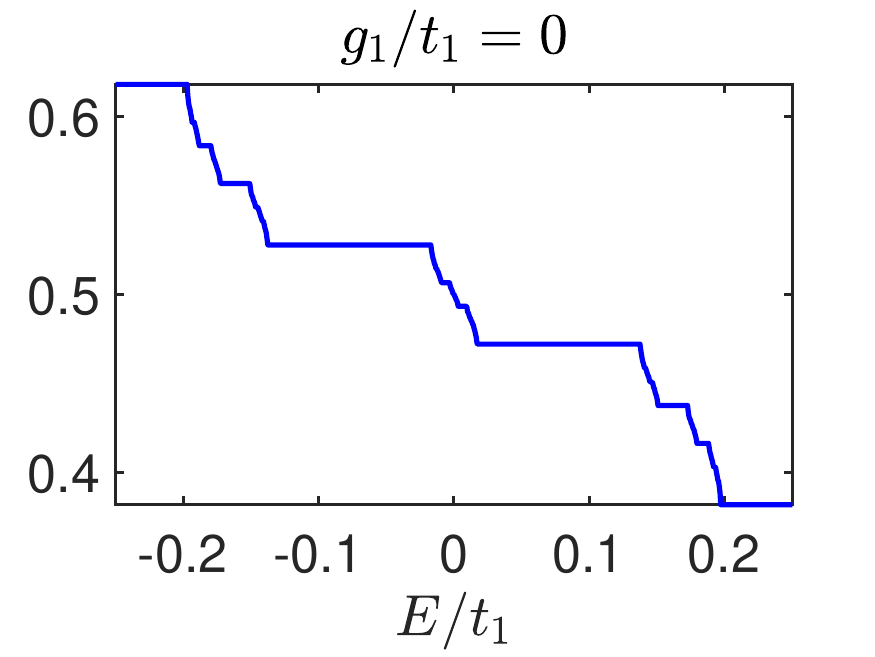} 
    }
    \caption{Spectral survival ratio $p(E)$ for $\beta =  (\sqrt5 -1)/2 $, $N=1597$, and   $(g_1, t_2,g_2) = (0,0,0)$. (b) Zoom-in view of Panel~(a).   
   }
    \label{Fig:p(E) and zoom_in-Appx}
\end{figure}

To illustrate the above features, we consider the spectral survival ratio $p(E)$ defined in Eq.~\eqref{Eq:percent_eig}, where plateaus in the $p(E)$ curves correspond to energy gaps in the spectrum. At the Aubry-Andr{\'e} criticality, the spectrum is known to resemble a Cantor set~\cite{Hofstadter:1976}, and the associated $p(E)$ curves for $\beta \approx 0.62$, shown in Fig.~\ref{Fig:p(E) and zoom_in-Appx}, naturally exhibit self-similar structures.
Specifically, the $p(E)$ curve can be divided into three segments, with the two largest gaps located at $p(E) = \beta$ and $p(E) = 1 - \beta$. Following the earlier discussion, we label the regions with $p(E) > \beta$, $1 - \beta < p(E) < \beta$, and $p(E) < 1 - \beta$ as the L, C, and R child spectra, respectively. Since all three segments share the same local variable as the parent, they exhibit self similarity.
Adopting the procedure in Ref.~\cite{Jensen:1983}, we ``flatten'' the $p(E)$ curve onto the energy axis and apply the box-counting method, allowing us to deduce a fractal dimension $d_{\rm H} \approx 0.54$, which agrees well with the SFD $\approx 0.52$ reported for the Aubry-Andr{\'e} spectrum~\cite{Falconer:1985}.

Next, we introduce the antisymmetric hopping term $g_1$ to examine its effect on the fractal structure of the spectrum. Two limiting cases serve as useful benchmarks:
(i) when $g_1 = 0$, the system exhibits clear self similarity at the critical point $V_c$;
(ii) when $g_1 = t_1$, self-similarity is absent at $V_c$ as the spectrum becomes gapless.
For intermediate values $0 < |g_1/t_1| < 1$, one thus expects a gradual suppression of self-similar features in the spectrum with increasing $g_1$.

\begin{figure}[t]
    \centering
    \stackinset{l}{0.1cm}{b}{3.cm}{\colorbox{white}{{}}}{
        \includegraphics[width=0.65\linewidth]
        {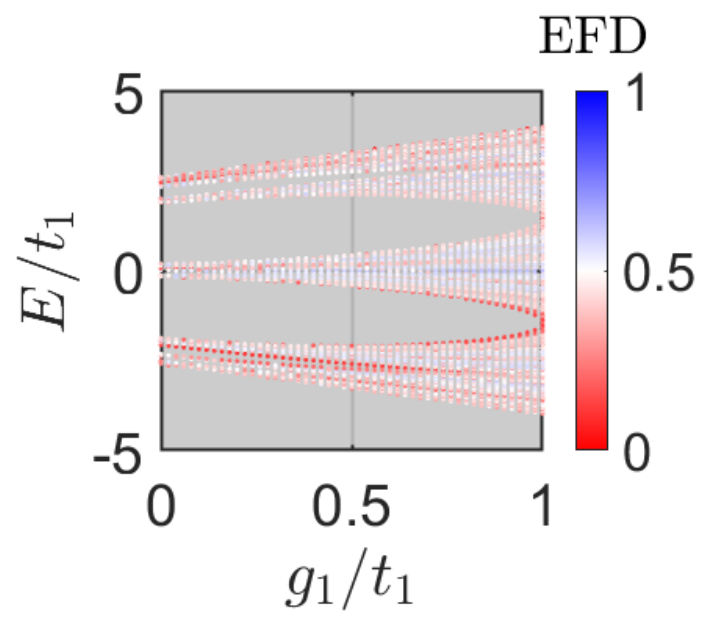}
    }    
    \caption{Antisymmetric hopping strength ($g_1$) dependence of PBC spectrum and EFD, where $V_0$ is chosen such that $V_0 = V_c (g_1)$ for the corresponding $g_1$ values. Gray background is used to enhance contrast with the color scheme of the data.
   }
    \label{Fig:spectrum_SFD_at_V_c}
\end{figure}

To further investigate this transition, we consider the Lebesgue measure (denoted by $\Delta W_c$) of the spectrum at the critical potential strength. As discussed in Ref.~\cite{Longhi:2021}, introducing a nonzero antisymmetric hopping $g_1$ renders the spectrum complex, and results in a finite Lebesgue measure in the thermodynamic limit ($N \to \infty$). In our notation, and for $t_1 \geqslant g_1 > 0$, the result from Ref.~\cite{Longhi:2021} can be expressed as
\begin{equation} \label{Eq:L_measure}
\Delta W_c = 4 \left\lvert (t_1 + g_1) - \sqrt{(t_1 + g_1)(t_1 - g_1)} \right\rvert,
\end{equation}
indicating that the quasiperiodicity-induced minibands acquire a finite bandwidth, i.e., a nonzero extent along the real energy axis.
As $g_1$ increases (within $|g_1| \leqslant |t_1|$), the Lebesgue measure grows, and the total gap size diminishes, leading to a suppression of the self-similar spectral structure\footnote{Strictly speaking, a spectrum with nonzero Lebesgue measure can no longer be a pure Cantor set. However, one cannot rule out the possibility of a mixed structure composed of continuous intervals (bands) and residual fractal features.}.
This trend directly correlates with the behavior of the SFD shown in Figs.~\ref{Fig:SFD-vs-V0_2}--\ref{Fig:SFD-vs-V0_3} of the main text, where increasing $g_1$ leads to a larger SFD at the critical potential strength.
In Fig.~\ref{Fig:spectrum_SFD_at_V_c}, we further present the PBC spectrum and the corresponding EFD. As $V_0$ increases, the energy gaps gradually shrink, resulting in a more continuous spectrum. This progression culminates at $g_1 = t_1$, where the SFD reaches unity, as shown in Fig.~\ref{Fig:SFD-vs-V0_3}(b).
Extending the above analysis, we also examine the spectral survival ratio for $g_1 \neq 0$, as shown in Fig.~\ref{Fig:p(E) and zoom_in}(b) in the main text.

\begin{figure*}[t]
   \centering 
    \stackinset{l}{-0.1cm}{b}{3.5cm}
    {\colorbox{white}{(a)}}{
        \includegraphics[width=0.315\linewidth]{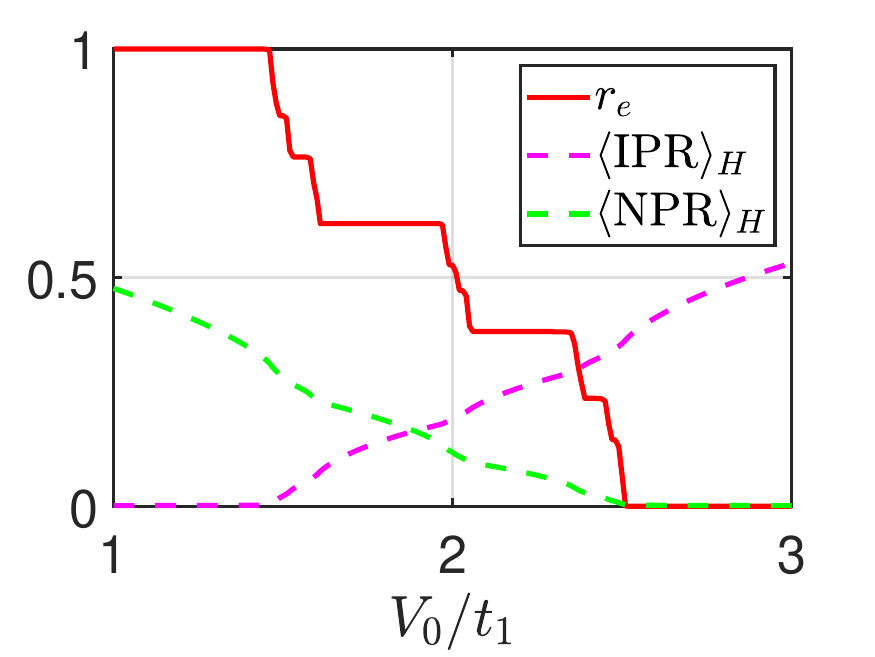}
        }
    \stackinset{l}{-0.4cm}{b}{3.5cm}
    {\colorbox{white}{(b)}}{
        \includegraphics[width=0.315\linewidth]{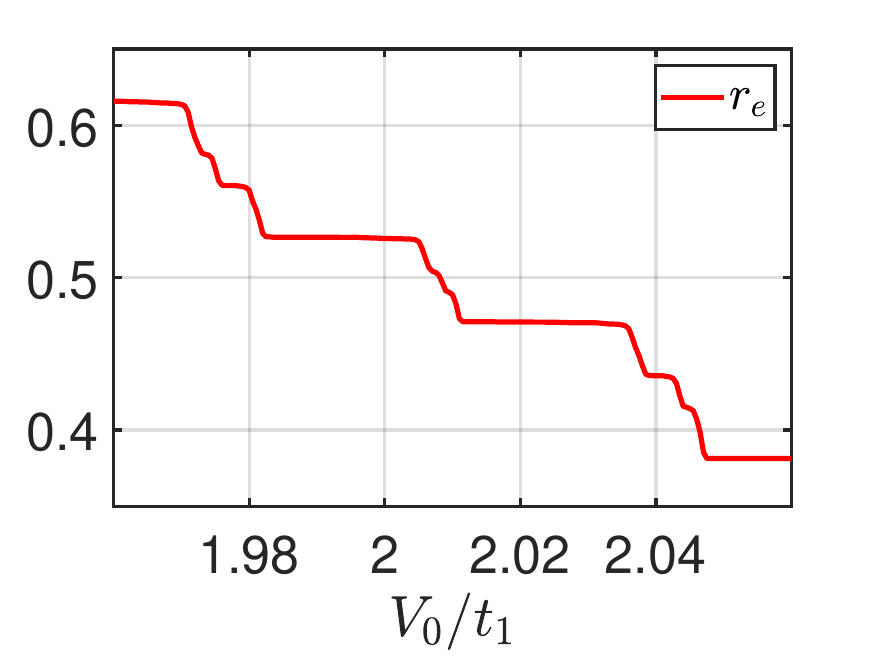}    
    }    
    \stackinset{l}{0cm}{b}{3.5cm}
    {\colorbox{white}{(c)}}{
        \includegraphics[width=0.315\linewidth]{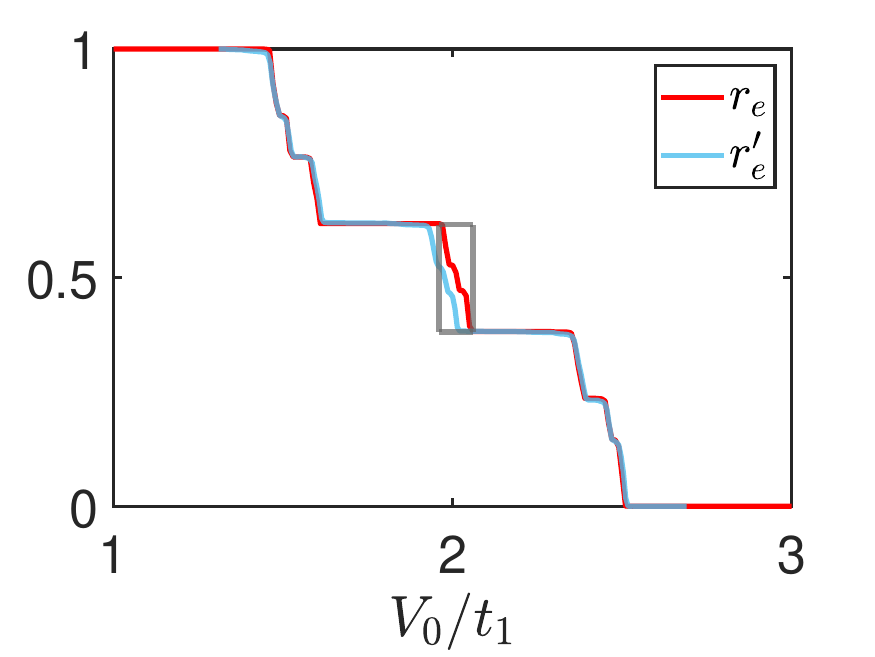}
    }  
    \caption{Similar plots to Fig.~\ref{Fig:re_IPR_NPR}, but in the Hermitian limit with $(g_1,t_2,g_2) = (0,0.1,0)$. 
    The other parameter values include (a) $N=1597$ and  $\delta_{V_0} = 10^{-2}$ and (b,c)  $N=15000$ and $\delta_{V_0} = 10^{-3}$. 
    }
    \label{Fig:t2_ME_rcre} 
\end{figure*}

\subsection{Systems with longer-range hopping terms}
 
Next, we include the longer-range hopping terms $t_2$ and $g_2$. These additional terms eliminate the critical potential strength and give rise to mobility edges, thereby enabling access to a regime where localized and extended states coexist.
In this regime, we systematically track changes in spectral characteristics (e.g., EER) as the potential strength $V_0$ varies.

\begin{figure}[t]
    \centering 
    \stackinset{l}{-0.1cm}{b}{3.65cm}{\colorbox{white}{(a)}}{
        \includegraphics[width=0.65\linewidth]{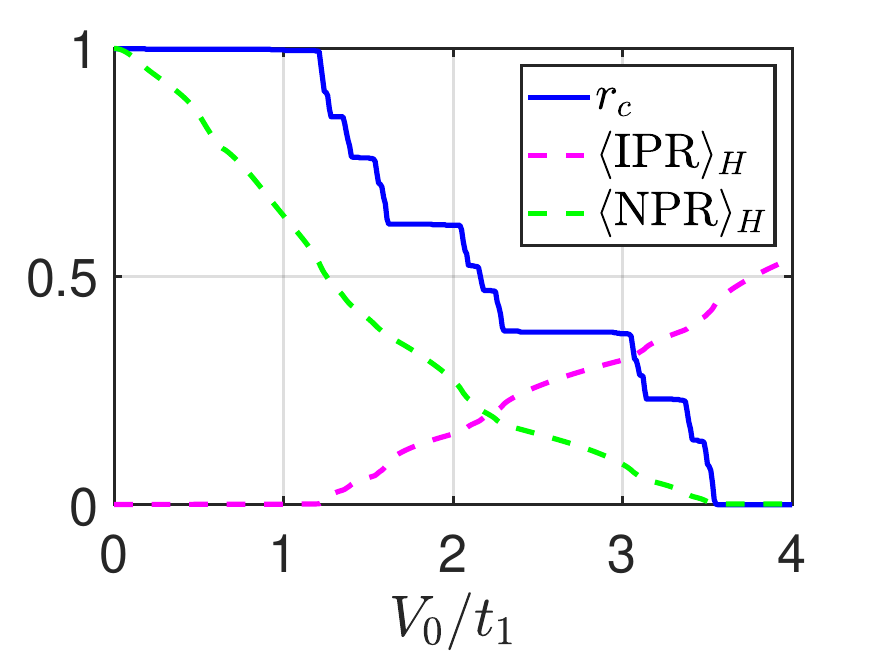}    
    } \\
    \stackinset{l}{-0.1cm}{b}{2.25cm}{\colorbox{white}{(b)}}{
        \includegraphics[width=0.45\linewidth]{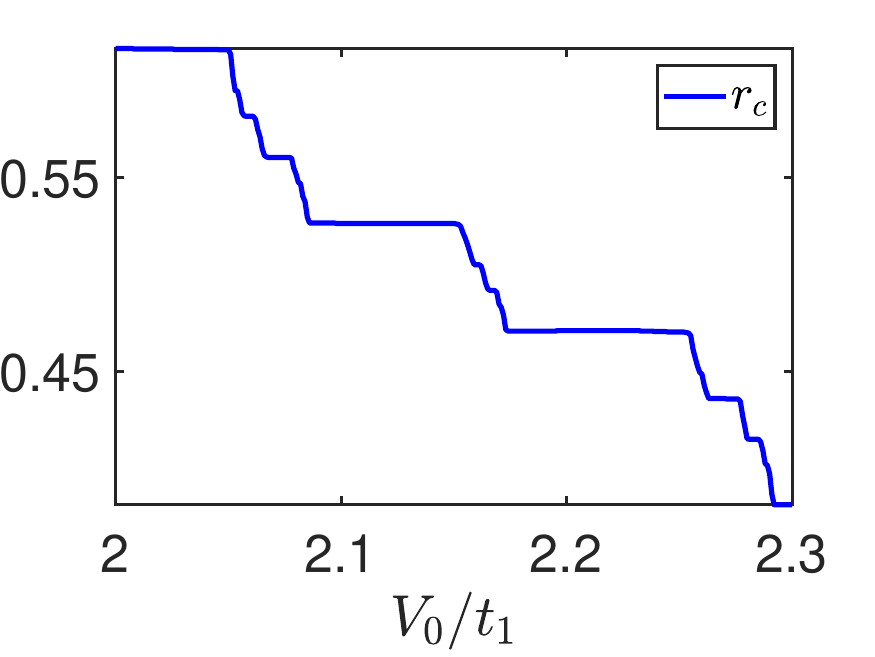}    
    }    
    \stackinset{l}{-0.05cm}{b}{2.35cm}{\colorbox{white}{(c)}}{
        \includegraphics[width=0.45\linewidth]{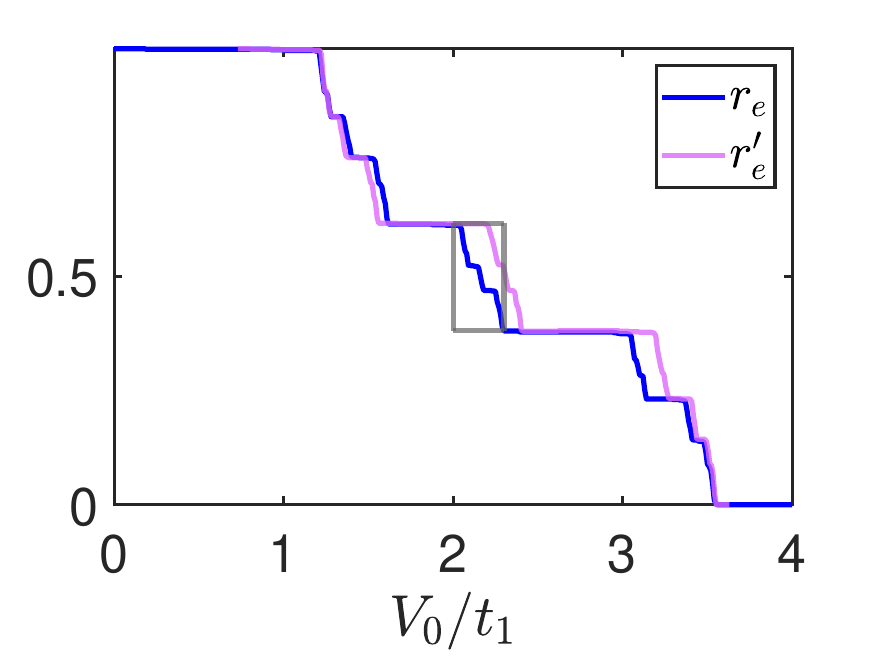}
    } \\
        \stackinset{l}{-0.1cm}{b}{2.3cm}{\colorbox{white}{(d)}}{
        \includegraphics[width=0.44\linewidth]{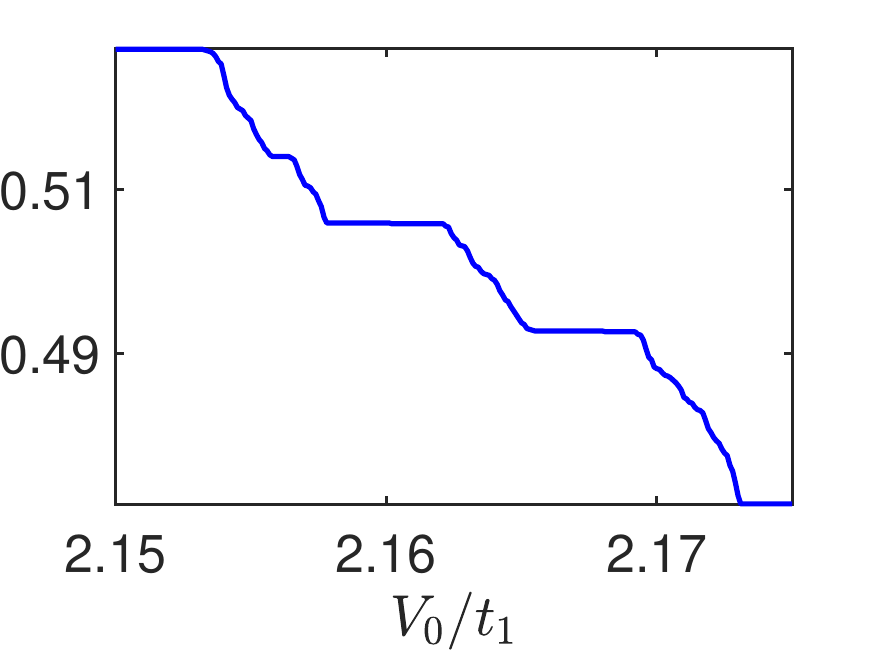}
    }
    \stackinset{l}{-0.05cm}{b}{2.3cm}{\colorbox{white}{(e)}}{
        \includegraphics[width=0.44\linewidth]{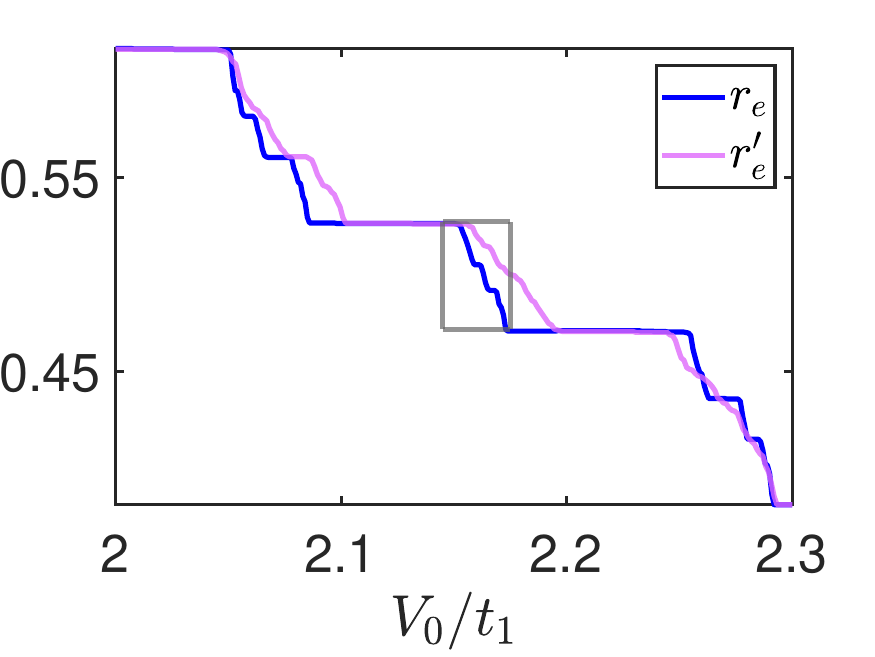}
    }    
    \caption{Similar plots to Fig.~\ref{Fig:rc_IPR_NPR}--\ref{Fig:rc_closer_view} but for a different parameter set, $(g_1,t_2,g_2) = (0.05,0.3,0)$. 
    The other parameter values include (a) $N=1597$ and  $\delta_{V_0} = 10^{-2}$, (b,c)  $N=8000$ and $\delta_{V_0} = 10^{-3}$, and (d,e)  $N=20000$ and $\delta_{V_0}= 10^{-4}$. 
    }
    \label{Fig:rc_IPR_NPR_old} 
\end{figure}

\begin{figure}[t]
    \centering 
    \stackinset{l}{0.2cm}{b}{3.9cm}{\colorbox{white}{(a)}}{
        \includegraphics[width=0.65\linewidth]{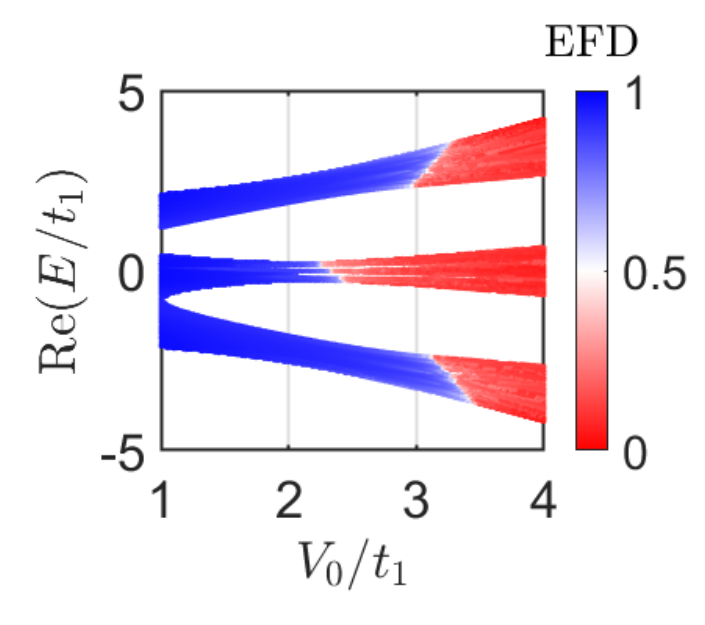}    
    } \\
    \stackinset{l}{-0.1cm}{b}{2.35cm}{\colorbox{white}{(b)}}{
        \includegraphics[width=0.45\linewidth]{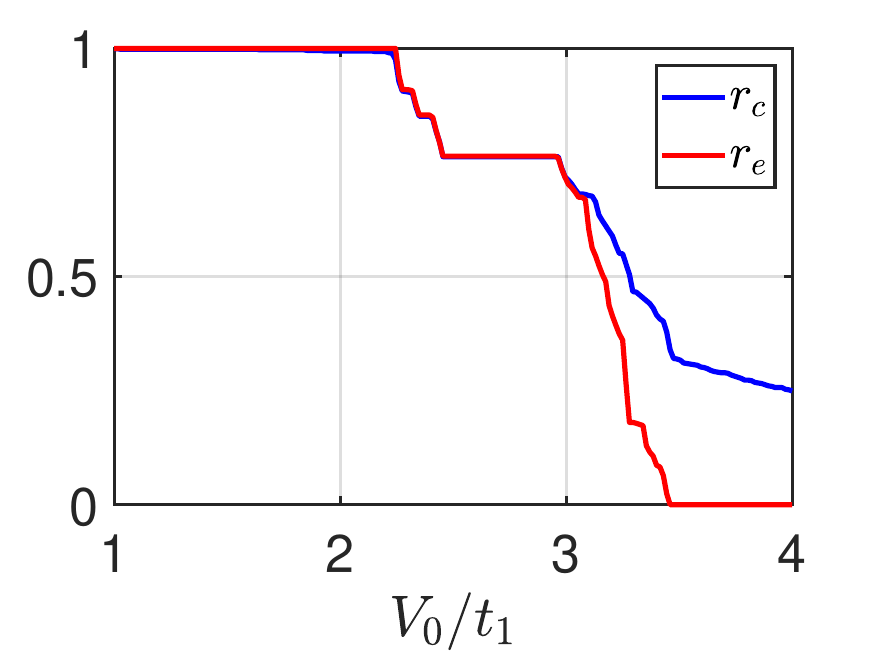}    
    }    
    \stackinset{l}{0cm}{b}{2.35cm}{\colorbox{white}{(c)}}{
        \includegraphics[width=0.45\linewidth]{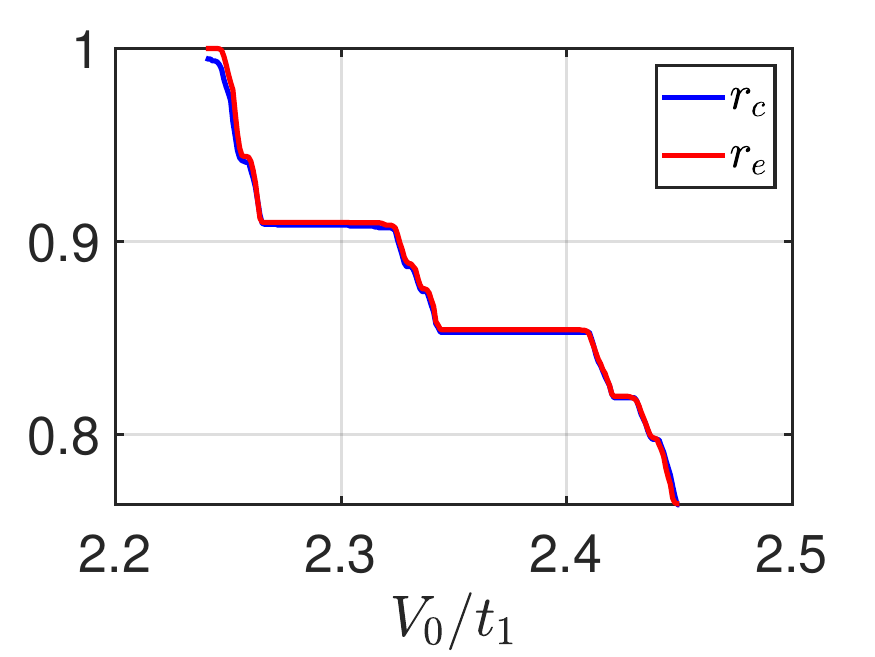}
    } 
    \caption{ EFD ($\Gamma_n$) and complex eigenvalue ratio ($r_c$), EER ($r_e$) with a larger $g_2$. 
    (a) $\Gamma_n$, and (b) $r_c$ and $r_e$ for  $\delta_{V_0} = 1.5 \times10^{-2}$ and $N= 1597 $. 
    (c) $r_c$ and $r_e$ for $\delta_{V_0} = 10^{-3}$ and $N=4181$.
    The adopted parameter values are given by $(g_1,t_2,g_2) = (0.05,0,0.3)$. 
        }
    \label{Fig:g2_ME_rcre} 
\end{figure}

For small but finite values of $t_2$ or $g_2$, the resulting mobility edge is expected to deviate only slightly from the critical value $V_c$ corresponding to $t_2 = g_2 = 0$, in the spirit of perturbation theory. It is therefore natural that the $r_e$ and $r_c$ plots in this regime, such as those in Figs.~\ref{Fig:re_IPR_NPR} and \ref{Fig:rc_IPR_NPR}, exhibit staircase-like structures.
As discussed in the main text and Appendix~\ref{Appx:r_c}, the inclusion of $t_2, g_2 \neq 0$ gives rise to plateau features in $r_e(V_0)$ and $r_c(V_0)$ that reveal self-similar patterns, reminiscent of the structures seen in $p(E)$ within the Hermitian regime. More quantitatively, self-similar behavior persists even when $t_2/t_1$ or $g_2/t_1$ are on the order of $0.1$,  
as demonstrated in Fig.~\ref{Fig:re_IPR_NPR} in the main text. Here, Fig.~\ref{Fig:t2_ME_rcre} additionally shows the self-similar behavior of $r_e$ in the Hermitian limit by setting $g_1$ to zero.
Empirically, numerical results show visible self-similar structures for $t_2/t_1 = 0.3$ (see Figs.~\ref{Fig:rc_IPR_NPR_old}) and even an unrealistically large value of  $g_2/t_1 = 0.3$ (see Fig.~\ref{Fig:g2_ME_rcre}).

While the inclusion of $g_2$ also induces a mobility edge, as shown in Fig.~\ref{Fig:g2_ME_rcre}(a), it exhibits a distinct shape from the effects of $t_2$: localization begins near the center of the spectrum.
Another notable feature is the deviation between $r_c$ and $r_e$ for $V_0 \geqslant 3.1$ [see Fig.~\ref{Fig:g2_ME_rcre}(b)]. As demonstrated in Figs.~\ref{Fig:g2_ME_rcre}(b,c), the self-similar structure for $g_2 \neq 0$ is more subtle and becomes visible only within a narrow range, $V_0 \in [2.2, 2.5]$, corresponding to the small steps in Fig.~\ref{Fig:g2_ME_rcre}(a).

In the regime of even larger $t_2$ or $g_2$, the system moves beyond the perturbative regime discussed earlier, and the self-similar structure generally disappears.  
As noted in the main text, this breakdown of self similarity is accompanied by the absence of a well-defined minimum in the SFD, as shown in Fig.~\ref{Fig:SFD-vs-V0_3}(a).

\subsection{Quasiperiodicity with a different $\lambda_0 / a_0$ ratio}

\begin{figure}[t]
    \centering
    \stackinset{l}{-0.2cm}{b}{2.7cm}{\colorbox{white}{(a)}}{
        \includegraphics[width=0.46\linewidth]{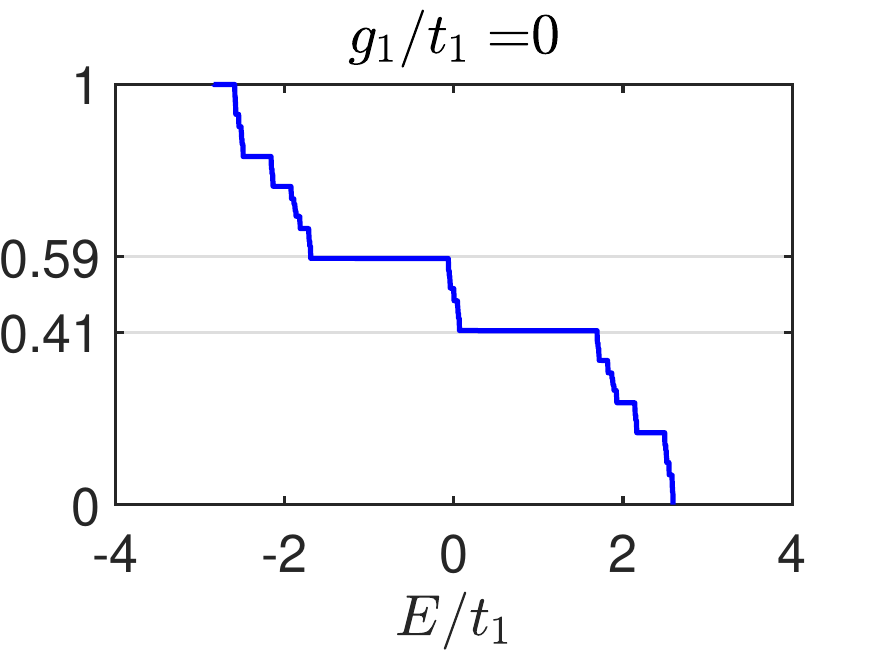}
    }
    \stackinset{l}{-0.2cm}{b}{2.7cm}{\colorbox{white}{(b)}}{
        \includegraphics[width=0.46\linewidth]{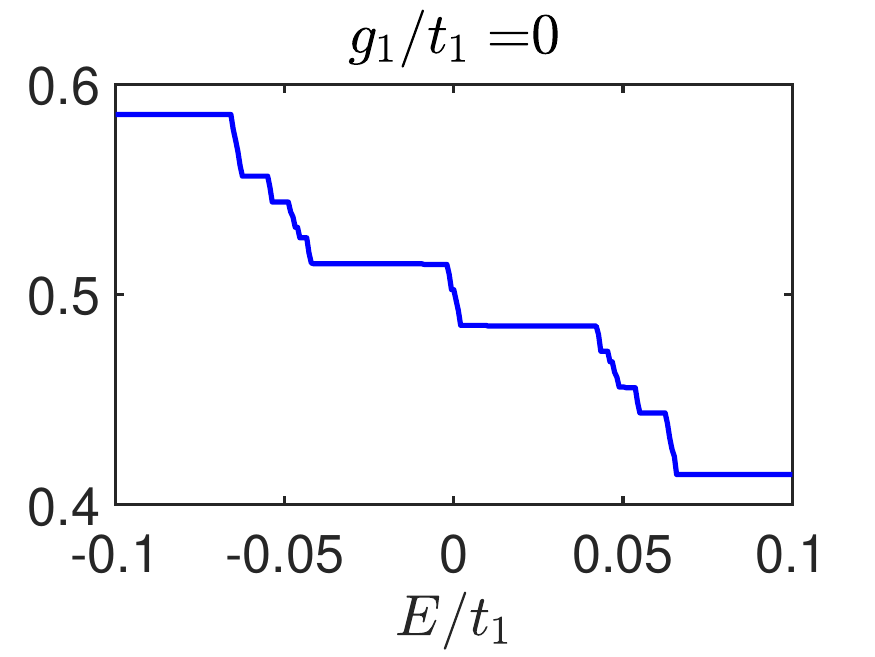} 
    }
    \stackinset{l}{-0.2cm}{b}{2.7cm}{\colorbox{white}{(c)}}{
        \includegraphics[width=0.46\linewidth]{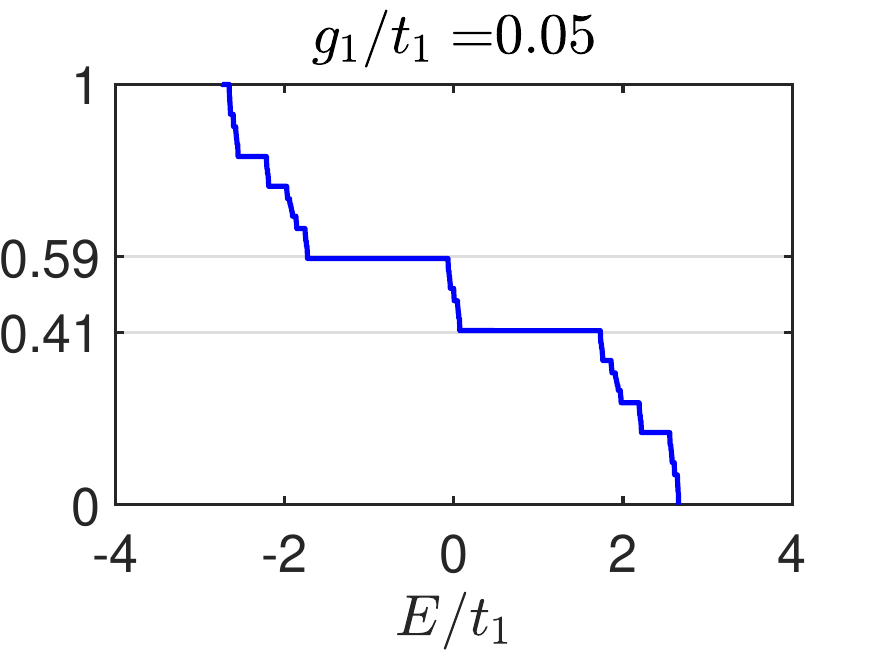}
    }
    \stackinset{l}{-0.2cm}{b}{2.7cm}{\colorbox{white}{(d)}}{
        \includegraphics[width=0.46\linewidth]{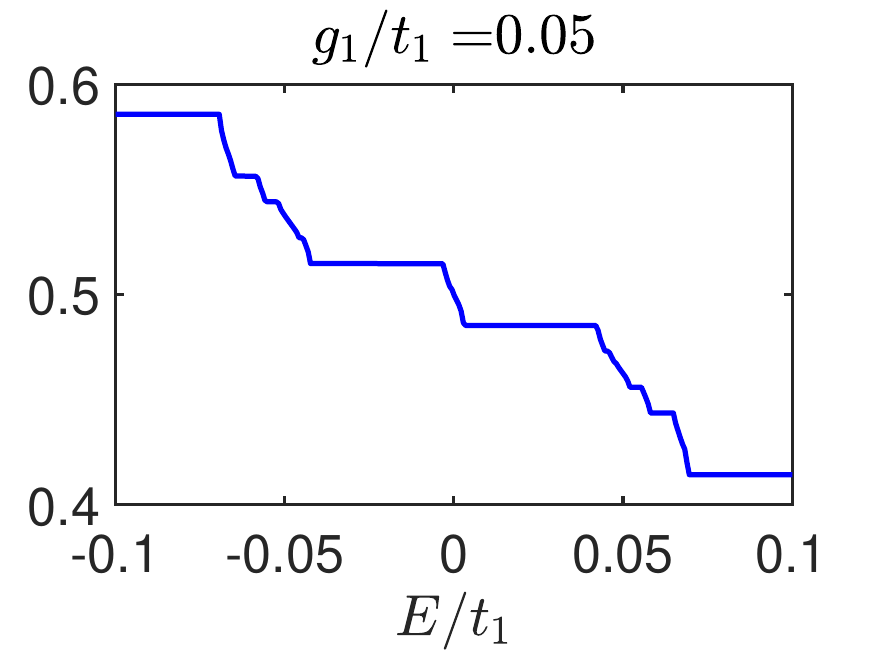} 
    }
    \caption{Spectral survival ratio $p(E)$ for $\lambda_0/a_0 = \sqrt{2} + 1$, corresponding to $\beta =  \sqrt{2} - 1 \approx 0.41$. (a) $p(E)$ for $N=3000$ and $(g_1, t_2,g_2) = (0,0,0)$. (b) Zoom-in view of Panel~(a). (c) Similar to Panel~(a) but for $(g_1, t_2,g_2) = (0.05,0,0)$. (d) Zoom-in view of Panel~(c), with $N=8000$.
   }
    \label{Fig:Silver_ratio}
\end{figure}

In this section, we examine the effects of the irrational period of the onsite potential.
When the quasiperiodicity is set by the silver ratio, $\lambda_0 / a_0 = \sqrt{2} + 1$, the corresponding local variable is $\beta = \sqrt{2} - 1$. 
The Hofstadter transformation~\cite{Hofstadter:1976} leads to
\begin{eqnarray}
    \alpha' &=& \frac{1}{\sqrt{2}-1} - \Big\lfloor\frac{1}{\sqrt2-1}\Big\rfloor = \beta, \\
    \beta' &=& {\rm Frac } \big[ (\frac{1}{\sqrt{2}-1}-2)^{-1} \big] =  \beta,
\end{eqnarray}
where the $\lfloor x \rfloor$ stands for the greatest integer less than or equal to $x$, and ${\rm Frac } [x]$ stands for the fractional part of $x$.
The above relation indicates that the child spectra L, R, and C all share the same arrangement of bands and gaps. As a result,  $p(E)$ is expected to exhibit behavior similar to the case of the golden ratio.
This expectation is confirmed numerically in Fig.~\ref{Fig:Silver_ratio}, which shows a self-similar staircase structure, both with and without $g_1$ term. As before, the two largest plateaus appear at $p(E) = \beta$ and $p(E) = 1 - \beta$.

\end{document}